\documentclass{article}

\usepackage{arxiv}
\usepackage[T1]{fontenc}    
\usepackage{hyperref}       
\usepackage{url}            
\usepackage{booktabs}       
\usepackage{amsfonts}       
\usepackage{microtype}      
\usepackage{graphicx}
\usepackage{natbib}
\usepackage{amssymb}
\usepackage{array}
\usepackage{threeparttable}
\usepackage{xcolor}
\usepackage{authblk}

\title{WeatherReal: A Benchmark Based on In-Situ Observations for Evaluating Weather Models}

\author[1]{\mbox{Weixin Jin}}
\author[1]{\mbox{Jonathan Weyn}}
\author[1]{\mbox{Pengcheng Zhao}}
\author[1]{\mbox{Siqi Xiang}}
\author[1]{\mbox{Jiang Bian}}
\author[1]{\mbox{Zuliang Fang}}
\author[*,1]{\mbox{Haiyu Dong}}
\author[1]{\mbox{Hongyu Sun}}
\author[1]{\mbox{Kit Thambiratnam}}
\author[1]{\mbox{Qi Zhang}}

\affil[1]{Microsoft Corporation}
\affil[*]{Corresponding to haiyu.dong@microsoft.com}

\date{}

\renewcommand{\headeright}{}

\begin{document}
\maketitle

\begin{abstract}
In recent years, AI-based weather forecasting models have matched or even outperformed numerical weather prediction systems. However, most of these models have been trained and evaluated on reanalysis datasets like ERA5. These datasets, being products of numerical models, often diverge substantially from actual observations in some crucial variables like near-surface temperature, wind, precipitation and clouds - parameters that hold significant public interest. To address this divergence, we introduce WeatherReal, a novel benchmark dataset for weather forecasting, derived from global near-surface in-situ observations. WeatherReal also features a publicly accessible quality control and evaluation framework. This paper details the sources and processing methodologies underlying the dataset, and further illustrates the advantage of in-situ observations in capturing hyper-local and extreme weather through comparative analyses and case studies. Using WeatherReal, we evaluated several data-driven models and compared them with leading numerical models. Our work aims to advance the AI-based weather forecasting research towards a more application-focused and operation-ready approach.
\end{abstract}

\section{Introduction}

Accurate weather forecasting plays a vital role in saving lives, aiding emergency management, and reducing the economic impact of severe weather events~\citep{bauer_quiet_2015}. The traditional paradigm of weather forecasting is numerical weather prediction (NWP), which focuses on nonlinear partial differential equations to simulate atmospheric dynamics and physical processes~\citep{benjamin_100_2019}. In recent years, with the advancement of artificial intelligence (AI) technology and the continuous accumulation of massive weather data, data-driven methods have been increasingly incorporated into various stages and different scales of weather forecasting~\citep{ravuri2021skilful, schultz2021can, weyn2021sub}. Particularly in the past two years, numerous data-driven models addressing the short to medium-range (0–10 day) forecasting problem have emerged~\citep{bi_accurate_2023, lam2023learning, chen2023fuxi, lang2024aifsecmwfsdatadriven}. These models have surpassed the operational Integrated Forecast System (IFS) from European Centre for Medium-Range Weather Forecasts (ECMWF) in metrics such as Root Mean Square Error (RMSE) and Anomaly Correlation Coefficient (ACC). These breakthroughs have instilled confidence that data-driven models can be significant tools for enhancing the accuracy and computational efficiency of weather forecasting.

It is important to note that long-term gridded reanalysis data, in particular ERA5~\citep{hersbach_era5_2020}, have greatly benefited the training and evaluation of these data-driven models. ERA5 is produced using 4D-Var data assimilation, which incorporates a wide range of direct and remote sensing observations, as well as model forecasts from the ECMWF IFS. ERA5 provides global gridded estimates at different atmospheric levels and near the surface with a resolution of 0.25° (approximately 30 km). Utilizing these gridded data, deep learning algorithms, such as Vision Transformers~\citep{pathak2022fourcastnetglobaldatadrivenhighresolution, bi_accurate_2023, chen2023fuxi} and Graph Neural Networks~\citep{keisler2022forecastingglobalweathergraph, lam2023learning, lang2024aifsecmwfsdatadriven}, can be easily applied as forecast models. WeatherBench2~\citep{rasp_weatherbench_2024} further provides a unified framework for these models with consistent training, ground truth, and baseline data, and offers standardized evaluation metrics, thereby facilitating easier direct comparisons between different models.

However, relying exclusively on reanalysis data for training and evaluating data-driven models, while simplifying the process, may also restrict their practical value in operational applications by introducing several risks. Firstly, reanalysis data inherently contains bias compared to actual weather conditions, if the initial conditions or background error covariances are not accurately represented during data assimilation. The simplification of physical processes and parameterizations for cloud formation, precipitation processes, and boundary layer dynamics may further impact variables such as clouds and precipitation~\citep{JIANG_ERA5_2021, lavers2022_ERA5, wu2023_era5}. Some studies also indicate that ERA5 has limited capability in representing extreme climate events such as cold snaps, heat waves, and heavy rainfall~\citep{tarek_2020_era5_eval, TAN_2023_era5_eval, ROFFE2023_era5_eval}. Secondly, although the resolution of reanalysis products has continuously improved, their near-surface variables fundamentally represent the average conditions within a grid interval, and thus will inevitably have some errors when applied to specific local positions. Finally, models trained on reanalysis data need to switch to analysis fields as initial conditions provided by numerical models during operation, which further impacts their skills. This issue is compounded by the design of the 12-hour assimilation window in ERA5 which includes information from the future and does not align with real-world conditions~\citep{hersbach_era5_2020}.In summary, building weather models based on reanalysis data can generally reflect the models’ abilities to capture temporal and spatial atmospheric variations, but some deficits are still present in assessing its performance for users and downstream applications.

Based on the aforementioned reasons, it is a common practice to correct reanalysis and NWP forecasts using raw observations~\citep{nwp_correction_gneiting2014calibration, era5_correction_essd-12-2097-2020}. However, given the robust learning capabilities of data-driven models, it is feasible to directly incorporate raw observations into model training and evaluation. Several studies have already dedicated to this direction. \cite{vaughan2024aardvark} developed an end-to-end data-driven weather forecasting system consisting of three modules: an encoder, a processor, and a decoder. The system utilizes in-situ observations and remote sensing data, allowing it to directly learn a mapping from raw input observations to output forecasts, and is evaluated using in-situ station observations. In another similar study~\citep{mcnally2024_ecobs}, near-surface station observation data is used diagnostically, meaning these data are predicted by the network but are not used as input. In the realm of short-term forecasting, ~\cite{andrychowicz2023deep} introduced MetNet3, which learns from both analysis and station observation data and makes forecasts up to 24 hours ahead for a range of weather parameters. These studies generally share some common issues: 

\begin{enumerate}
  \item Due to the complexity of raw observational networks, collecting and processing global raw observations is an extremely burdensome task. The datasets used in these studies are either concentrated in specific regions~\citep{andrychowicz2023deep}, come from a single data source with limited number of stations~\citep{vaughan2024aardvark}, or are even simulated by randomly masking reanalysis data~\citep{xiao2023fengwu, huang2024diffda}. This limits the value of the results.
  \item Although reanalysis data contains certain bias, numerical models effectively ensure its fidelity. In contrast, raw observations often exhibit numerous errors due to sensor and module malfunctions. To ensure the high quality of observational data, a well-designed quality control system is vital~\citep{wmo_guide_2010}. However, the datasets used in current studies often lack uniform and rigorous quality control.
  \item Due to the lack of widely recognized and readily available datasets like ERA5 for raw observations, different studies train and evaluate models based on different datasets and standards. This makes it challenging to compare the capabilities of these models.
\end{enumerate}

To address the above issues, we introduce WeatherReal, which comprises three near-surface observational datasets, two from diverse sources covering worldwide regions and one from weather reports collected from MSN weather users; a meticulously designed quality control system for raw observational data; and a benchmark for evaluating weather models, providing a unified standard to compare different models based on in-situ observations. The primary contribution of this work is the provision of a unified, reliable, and easily accessible benchmark for the study and application of weather models. Given the rapid advancement in data-driven weather models, Comparing different models on "real" observations from near-surface direct measurements, as a starting point, underscores the growing emphasis on the practical value of these models to people’s daily lives.

In Section 2, we provide an overview of the WeatherReal dataset. Section 3 outlines the quality control algorithms implemented for WeatherReal. In Section 4, we present the basic performance and benefits of WeatherReal through statistical analysis and case studies. Finally, Section 5 and 6 demonstrate the use of the benchmark by evaluating various types of forecasts using WeatherReal, and propose a few possible tasks with specific scoring methods, respectively. 

\section{Datasets}

WeatherReal includes three versions of datasets, two of which are derived from global near-surface in-situ observations: WeatherReal-ISD, An observational dataset based on the publicly available Integrated Surface Database (ISD), leveraging data from high-quality observation networks and has been subjected to rigorous post-processing and quality control through our independently developed algorithms; WeatherReal-Synoptic, An observational dataset from Synoptic Data PBC, a data service platform for 150,000+ in-situ surface weather stations, offering a much more densely distributed network. The third dataset is derived from weather reports collected from MSN weather users, directly capturing user-perceived weather conditions. Although currently unavailable, it is under development and expected to be released in a future version. Meanwhile, all the algorithms discussed in this paper, along with the WeatherReal-ISD dataset in 2023, will be publicly available on the official \href{https://github.com/microsoft/WeatherReal-Benchmark}{GitHub project page}. The variables and related information of these datasets are introduced in Table~\ref{tab:variables}.

\begin{table}[htbp]
\caption{Variable information in WeatherReal}
\label{tab:variables}
\centering
\begin{threeparttable}
\begin{tabular}{@{}lllll@{}}
\toprule
Variable & Short Name & Unit\textsuperscript{1} & Source column in ISD & Source subset in Synoptic \\
\midrule
2m Temperature & t & °C & TMP & air\_temp\_set\_1 \\
2m Dewpoint Temperature & td & °C & DEW & dew\_point\_temperature\_set\_1d \\
Surface Pressure\textsuperscript{2} & sp & hPa & MA & pressure\_set\_1d \\
Mean Sea-level Pressure & msl & hPa & SLP & sea\_level\_pressure\_set\_1d \\
10m Wind Speed & ws & m/s & WND & wind\_speed\_set\_1 \\
10m Wind Direction & wd & degree\textsuperscript{3} & WND & wind\_direction\_set\_1 \\
Total Cloud Cover & c & okta\textsuperscript{4} & GA, GD, GF, GG & Not used \\
1-hour Precipitation & ra1 & mm & AA & precip\_accum\_one\_hour\_set\_1 \\
3-hour Precipitation & ra3 & mm & AA & precip\_accum\_three\_hour\_set\_1 \\
6-hour Precipitation & ra6 & mm & AA & precip\_accum\_six\_hour\_set\_1 \\
12-hour Precipitation & ra12 & mm & AA & Not used \\
24-hour Precipitation & ra24 & mm & AA & precip\_accum\_24\_hour\_set\_1 \\
\bottomrule
\end{tabular}
\begin{tablenotes}
\item[1] Refers to the units used in the WeatherReal-ISD data. For the units provided by the raw ISD and Synoptic, please consult their respective documentation.
\item[2] For in-situ weather stations, surface pressure is measured at the sensor’s height, typically 2 meters above ground level at the weather station.
\item[3] Wind direction is measured clockwise from true north, ranging from 1° (north-northeast) to 360° (north), with 0° indicating calm winds.
\item[4] Okta is a unit of measurement used to describe the amount of cloud cover, with values ranging from 0 (clear sky) to 8 (completely overcast).
\end{tablenotes}
\end{threeparttable}
\end{table}

\subsection{WeatherReal-ISD}

The data source of WeatherReal-ISD, ISD~\citep{smith_integrated_2011}, is a global near-surface observation dataset compiled by the National Centers for Environmental Information (NCEI). More than 100 original data sources, including SYNOP (surface synoptic observations) and METAR (meteorological aerodrome report) weather reports, are incorporated.

There are currently more than 14,000 active reporting stations in ISD and it already includes the majority of known station observation data, making it an ideal data source for WeatherReal. However, the observational data have only undergone basic quality control, resulting in numerous erroneous data points. Therefore, to improve data fidelity, we performed extensive post-processing on it, including station selection and merging, and comprehensive quality control. These processes are described in Section~\ref{sec:isd_advanced}.

\subsection{WeatherReal-Synoptic}
\label{section:synoptic}

Data of WeatherReal-Synoptic is obtained from Synoptic Data PBC, which brings together observation data from hundreds of public and private station networks worldwide, providing a comprehensive and accessible data service platform for critical environmental information. For further details, please refer to \href{https://synopticdata.com/solutions/ai-ml-weather/}{Synoptic Data's official site}. The WeatherReal-Synoptic dataset utilized in this paper was retrieved in real-time from their Time Series API services in 2023 to address our operational requirements (see the last column of Table~\ref{tab:variables}), and the same data is available from them as a historical dataset. For precipitation, Synoptic also supports an advanced API that allows data retrieval through custom accumulation and interval windows, which is not covered in this paper. WeatherReal-Synoptic encompasses a greater volume of data, a more extensive observation network, and a larger number of stations compared to ISD. Note that Synoptic provides a quality control system as an additional attribute alongside the data from their API services, thus the quality control algorithm we developed independently has not been applied to the WeatherReal-Synoptic dataset.

\subsection{User Reports from MSN Weather}

In addition to the above two datasets, this paper incorporates user report data from \href{https://msn.com/weather}{MSN Weather}. MSN Weather is a comprehensive global weather forecasting service integrated within the Microsoft Start ecosystem. It offers real-time weather data and detailed forecasts, leveraging advanced AI-driven techniques. In order to further improve the quality of weather forecasts, it continuously collects a large number of user reports through its weather service. Users can report the weather at their location, including the current approximate temperature (whether higher or lower than the forecast value), sky condition (clear, cloudy, or overcast), and whether there is precipitation along with other weather phenomena.

These user reports provide an invaluable complement to traditional station observations by directly reflecting individual perceptions of the weather, and providing more immediate feedback. This unique dataset enables the calibration and validation of forecast models against real-world conditions, ensuring that the models more closely align with the public's day-to-day experiences, rather than with the measurements from sensing equipment under standardized scenarios.

However, the systematic integration of user reports requires extended validation, data accumulation, and the development of more sophisticated quality control algorithms. Effectively addressing the variability introduced by subjectivity in user reports presents a significant challenge. In this intitial version of WeatherReal, we choose to only demonstrate the unique value of user reports through case studies presented in the paper. A public release of validated user-reported data will be available in the near future.

\section{\label{sec:isd_advanced}Advanced Data Processing}

Raw in-situ observations require consistent and thorough data calibration. Since WeatherReal-ISD is directly derived from raw data compiled by the NCEI, we use it to introduce our data processing workflow. First, the data is extracted and resampled to an hourly level, which is the highest temporal resolution for most stations. Due to the complexity of data sources and changes in station IDs, some stations are merged based on metadata and data similarity. Finally, a set of quality control algorithms are conducted on the hourly data to detect and eliminate erroneous values.

\subsection{Data Extraction}

Different variables are first extracted from the corresponding columns in the raw ISD data (see Table~\ref{tab:variables}). Specifically, for total cloud cover, all related columns representing cloud cover at different layers are extracted, and the maximum value among them is taken as the total cloud cover. For precipitation, records are extracted if the time period is 1, 3, 6, 12, or 24 hours. For records with missing wind direction but zero wind speed, the wind direction is set to zero to represent calm wind.

The raw ISD data includes its own quality control flags. To maximize data retention, only values explicitly identified as erroneous are excluded from the dataset.

Timestamps in the raw ISD records are typically accurate to the hour or minute. The records are first rounded to the nearest hour. For any single hour with multiple entries, the value closest to the top of the hour is selected. This step is applied separately to each variable. However, for temperature and dewpoint temperature, the selected values should come from the same record. Therefore, records where both variables are available are given priority; if only one of them is available, temperature is preferred. For wind speed and wind direction, only records which have both of them are selected.

\subsection{Station Merging}

While the ISD incorporates a comprehensive 11-digit ID system, different IDs may actually represent the same station with essentially consistent data, and stations may also change IDs over time~\citep{dunn_hadisd_2012}. Given the inherently uneven distribution of stations, duplicate station data within a local area could further skew the weighting of model evaluation results for different regions. Therefore, IDs that most likely represent the same station are merged.

Merge candidates are initially selected based on their metadata, including latitude, longitude, elevation, and station name, following a similar approach to ~\cite{rennie_international_2014}. In this approach, we compute a similarity index consisting of three components: the geographical distance based on latitude and longitude, following an exponential decay with an e-folding distance of 25 km; the elevation difference, with an e-folding decay of 100 m; and the Jaccard similarity index~\citep{jaccard_1901} of the station names. The weighted sum of the three components is used as the final criterion, with weights consistent with ~\cite{rennie_international_2014}, and an acceptance threshold of 0.3. Additionally, since ISD stations include three sets of ID systems (USAF, WBAN, and CALL), stations with the same ID in any of these systems will also be considered as candidates.

\begin{figure}[htbp]
    \centering
    \includegraphics[width=\linewidth]{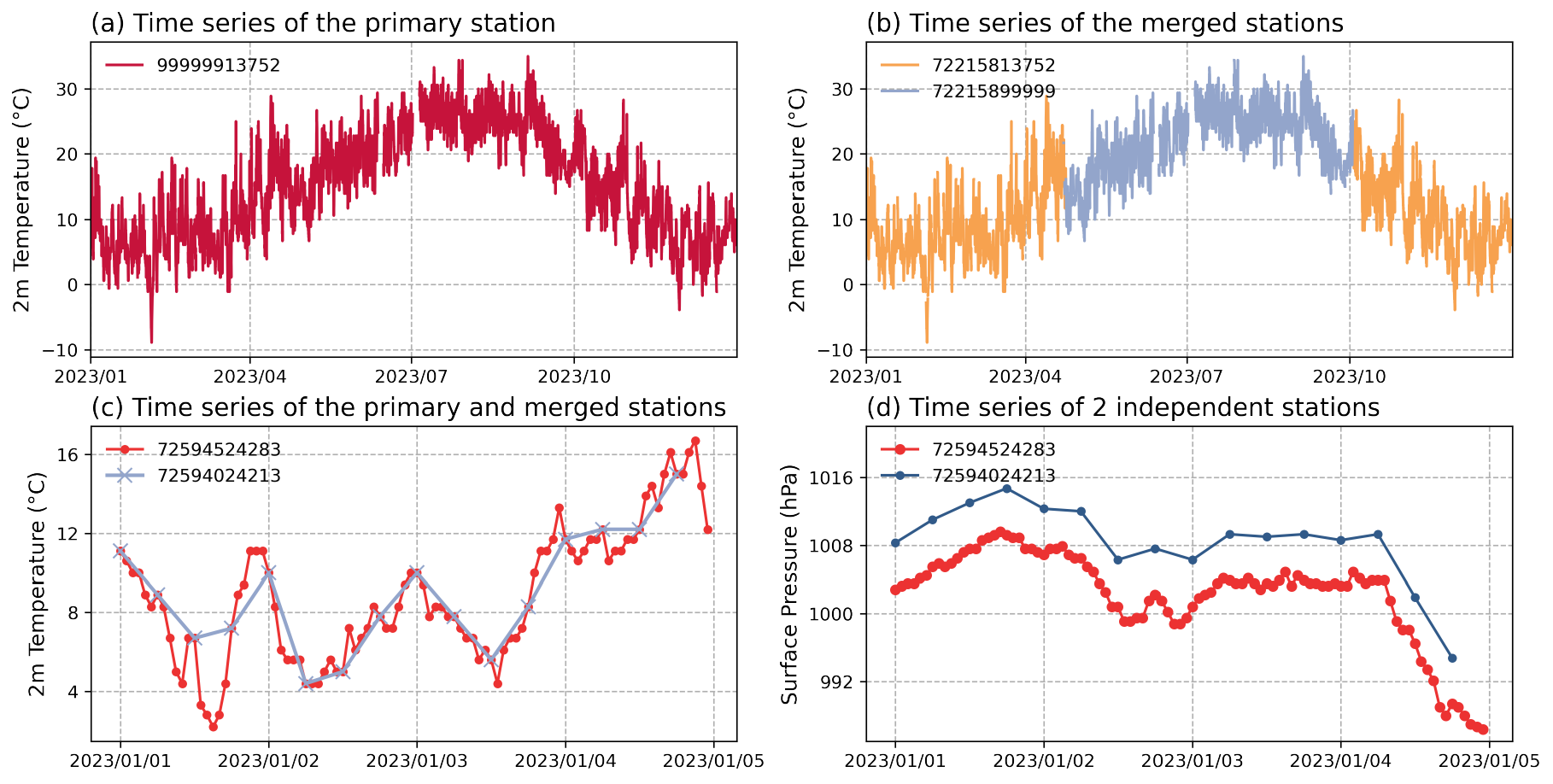}
    \caption{Example time series of primary and merged stations. (a) 2m temperature of the primary station 99999913752 and (b) the merged stations 72215813752 and 7221589999 in 2023. (c) 2m temperature and (d) surface pressure of stations 72594524283 and 72594024213 in 2023.}
    \label{fig:station_merging}
\end{figure}

Station pair candidates that pass the metadata similarity check are further compared based on their observation values. Stations are merged if at least 70\% of their simultaneous observation values are consistent. Missing values in the station with more available data points are filled using data from the other station. Figure ~\ref{fig:station_merging}a-b illustrates an example of three stations being merged. The primary station ID, 99999913752, recorded hourly temperature in 2023. A secondary station ID, 72214813752, had nearly identical records but the data was missing from April 23 to October 2, 2023. A third station ID, 72215899999, had records that complemented this gap. Therefore, these two secondary stations were merged into the primary station.

Station merging is performed on a per-variable basis as it is found that stations may have identical values for certain variables while showing discrepancies in others. The primary station 72594524283 recorded hourly temperature in 2023, while the secondary station 72594024213 recorded almost identical values at a lower frequency (Figure ~\ref{fig:station_merging}c). However, the surface pressures from these two stations were different (Figure ~\ref{fig:station_merging}d), thus their surface pressures are not merged. In the final dataset, all temperature data is excluded but surface pressure data is retained at station 72594024213.

\subsection{Quality Control}
\label{section:qc}

A fully automated quality control process is implemented to ensure the reliability of the dataset. Different quality control algorithms are applied to various ranges of variables (Table~\ref{tab:qc_summary}). These checks, which examine the value range, spatio-temporal variations, and inter-variable correlations, aim to remove outliers and in-homogeneous data points. The implementation for each algorithm is detailed below.

\subsubsection{Value Range Check}

\textbf{Known Records.} Absolute limits are first assigned to remove all records that fall outside the specified range. These limits are based on the global extreme values as reported by the World Meteorological Organization (WMO)~\citep{WMO_record_2024}. For station pressure, which depends on the elevation of the station, we use a range of 300-1100 hPa. Maximum 3-hour and 6-hour precipitation are also not specified by ~\cite{WMO_record_2024}. For maximum 3-hour precipitation, we set the upper limit to three times the 1-hour precipitation record extreme, based on the assumption that each hour in the 3-hour period should not exceed the 1-hour record extreme. For 6-hour precipitation, we set the upper limit to the 12-hour precipitation record extreme, as the 6-hour amount should logically not exceed the 12-hour extreme value. Additionally, this step includes logical absolute limits, such as non-negative values for precipitation and wind speed, a wind direction range of 0-360, and a total cloud cover range of 0-8.

\begin{figure}[htbp]
    \centering
    \includegraphics[width=\linewidth]{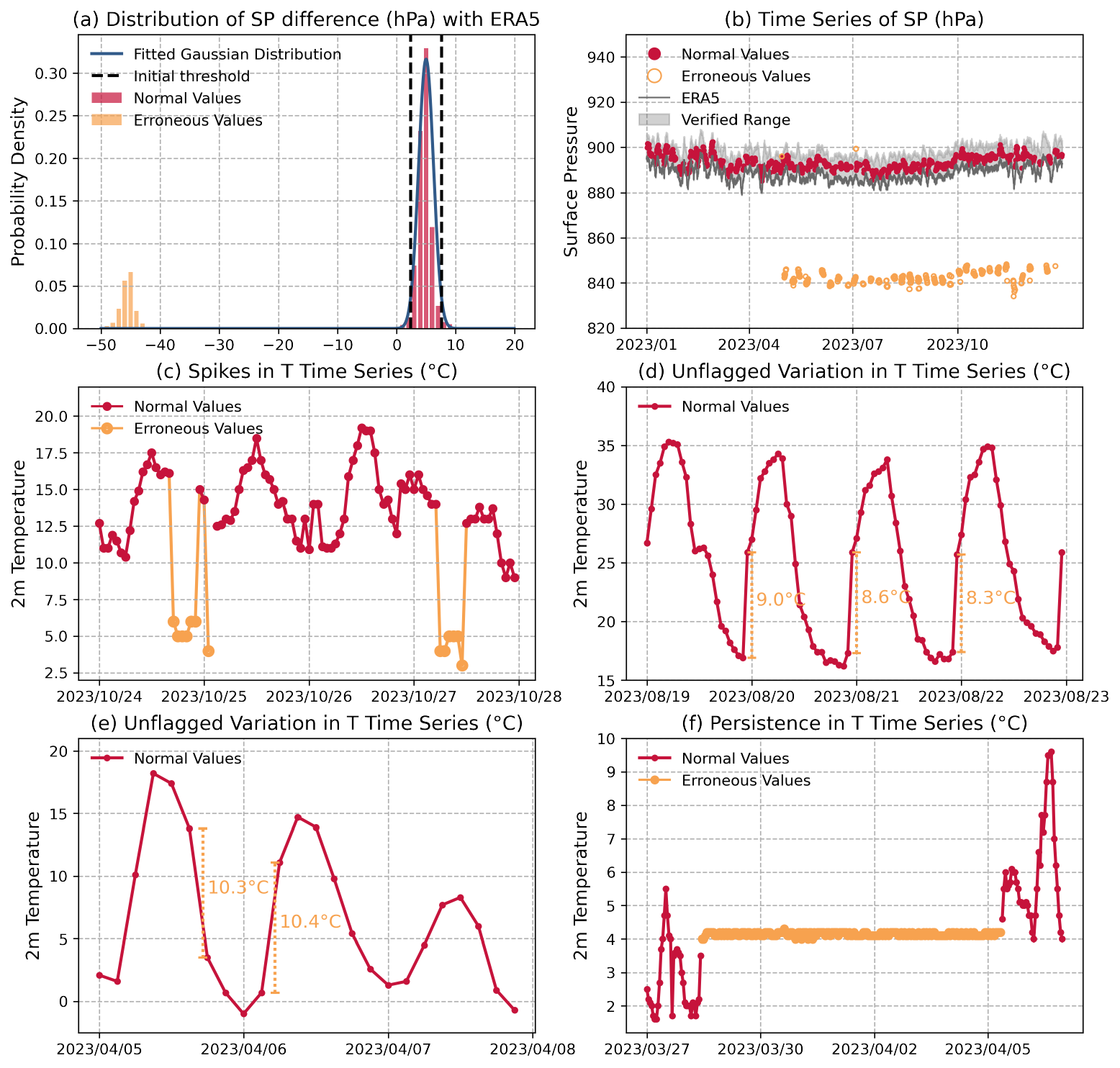}
    \caption{(a) Distribution of surface pressure differences between station observation and ERA5 at station 40310099999 in 2023, along with Gaussian fitting, and (b) the raw time series from the station and ERA5. The gray shaded area indicates the range within which the station observations will pass the quality control checks. (c) Identified spike errors in 2m temperature at station 12822099999 in 2023. (d) Unflagged spike in 2m temperature at station 94105099999 in 2023. (e) Spike unset by diurnal cycle check in 2m temperature at station 28612099999 in 2023. (f) Identified persistence errors in 2m temperature at station 06138099999 in 2023.}
    \label{fig:QC_step_check}
\end{figure}

\textbf{Distributional Gap.} A common quality control method involves comparing observed data with the historical data of the station. However, to retain as many stations as possible, WeatherReal provides datasets on an annual basis without considering the length of the station's historical data. To more accurately identify outliers, WeatherReal is compared with ERA5. For temperature, dewpoint temperature, surface pressure, and mean sea-level pressure, the differences between the two datasets generally follow a Gaussian distribution. A Gaussian curve is fitted to the difference series over one year at each station to detect outliers. To minimize the impact of outliers on fitting, the mean of the Gaussian distribution is set to the median of the difference series, and the standard deviation is set to 1.48 times the MAD (Median Absolute Deviation), based on the properties of the Gaussian distribution. All records falling outside the 0.01 probability density range, that is, those data points where the discrepancies between station observation and ERA5 data are abnormally large, are sorted. A value-by-value outward search is performed from the position where the probability density is 0.01, to find a gap exceeding a multiple of the MAD. If such a gap is found, all values outside this gap are flagged. This way, the unique information of the station observations compared to ERA5 is preserved, while outlying values are flagged. Figure~\ref{fig:QC_step_check}a-b show the surface pressure at station 40310099999 in 2023. The station occasionally recorded significantly lower values compared to other periods and ERA5. The Gaussian fitting considered only the data concentrated between 880-900 hPa, while data below 860 hPa were flagged as erroneous.

\begin{figure}[htbp]
    \centering
    \includegraphics[width=\linewidth]{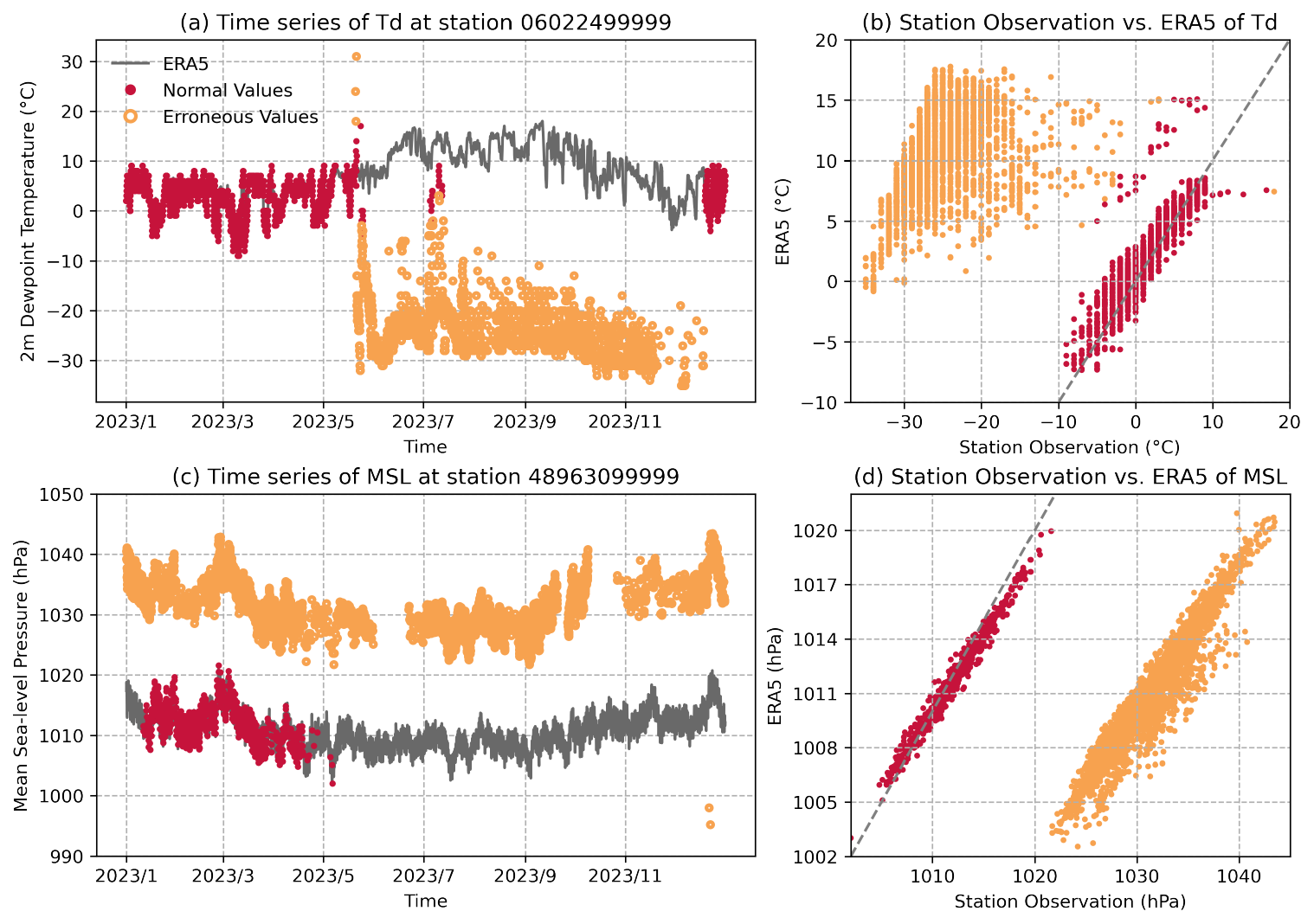}
    \caption{(a) Time series of dew point temperature observation and ERA5 at station 06022499999 in 2023 and (b) their scatter plot with the x-axis representing station observation and the y-axis representing ERA5. (c) Time series of mean sea-level pressure observations and ERA5 at station 48963099999 in 2023 and (d) their scatter plot. The grey dashed lines in (b) and (d) mark the diagonal of equality between station observation and ERA5.}
    \label{fig:dbscan}
\end{figure}

\textbf{Cluster Deviation.} Although the distributional gap method employs the median and MAD to fit the observation, thereby reducing the influence of outliers, its effectiveness diminishes when outliers constitute more than half of the data. This issue frequently arises in dewpoint temperature. Station 06022499999 in 2023 showed long-term anomalies from May to December (Figure~\ref{fig:dbscan}a), likely due to broken observation instruments, resulting in excessively high median and MAD. To resolve these issues, DBSCAN (Density-Based Spatial Clustering of Applications with Noise~\citep{ester1996density}) is applied to assess the differences between the station observation and ERA5. DBSCAN clusters the data based on data density, eliminating sparse noise. When multiple clusters are detected, only the cluster closest to ERA5 is retained, even if it contains fewer data points (Figure~\ref{fig:dbscan}b). This step is also critical for pressure variables. For the issue illustrated in Figure~\ref{fig:QC_step_check}a, when outliers are in the majority, the distributional gap check will have a counterproductive effect. As an example (Figure~\ref{fig:dbscan}c-d), normal values in the mean sea-level pressure observation at station 48963099999 in 2023, are concentrated during Jan to Apr 2023, and only accounted for 15\% of the total available data. Through DBSCAN, the abnormally higher values which might come from a wrong data source are removed. The median and MAD in the distributional gap check will be calculated only from normal values (the red points).

\subsubsection{Time Series Check}

\textbf{Spike.} Errors in observations are often accompanied by abrupt changes in values. A set of fixed thresholds is used to detect abnormal rate of change within a 3-hour period. When a record differs from the previous one by more than the threshold, both records are flagged. The algorithm then performs a bidirectional search to determine the end of the spike. This search continues until the value jumps in the opposite direction by more than the threshold, and the new value is close to the record preceding the spike (with a difference less than half the initial spike change). All intermediate values are flagged. Figure~\ref{fig:QC_step_check}c illustrates an example of a flagged spike in temperature. If the values gradually approach the record preceding the spike's start without abrupt changes exceeding the threshold (a common occurrence at stations with significant diurnal temperature variations, see Figure~\ref{fig:QC_step_check}d), no additional values are flagged. 

\textbf{Persistence.} A common observational error occurs when a station records the same value continuously over an extended period. When all observed values are nearly identical (e.g., the same value recorded continuously for 24 hours, or the same value recorded every 3 hours for 48 hours), all such values are flagged. Specifically, values of zero in precipitation and wind, as well as values of zero and eight oktas in cloud are excluded from this check. In practice, the standard deviation of the values within a variable-length sliding window is calculated. If it is below an acceptable minimum, all data within the window are flagged. This approach allows for the detection of persistence despite minor fluctuations (Figure~\ref{fig:QC_step_check}f).

\subsubsection{Cross-Variable Check}

\textbf{Supersaturation.} Dewpoint temperature is the temperature to which air must be cooled to become saturated with water vapor. Since this results in condensation and there are usually ample "nuclei" for condensation in the form of dew at the surface, it is very unusual for a station to record a super-saturation observation where the dewpoint temperature is higher than the air temperature. For any record where this occurs, both parameters are removed.

\textbf{Wind Consistency.} Since the range of wind direction in the ISD is from 1 to 360 degrees, with a zero value indicating calm winds, if either wind speed or wind direction is zero while the other is non-zero, both values are removed.

\textbf{Precipitation Consistency.} Logically, the accumulated precipitation over a longer period should not be less than that over a shorter period at the same timestamp. If these values contradict each other, both are removed. Notably, in this dataset, when the longer-term precipitation is zero, all shorter-term precipitation values for the fully-covered time periods are also set to zero. For instance, if the 6-hour precipitation at 18:00 is zero, then the 1-hour precipitation from 13:00 to 18:00 and the 3-hour precipitation from 15:00 to 18:00 are both set to zero. Conversely, if consecutive shorter-term precipitation values can be aggregated to form longer-term precipitation values (e.g., combining the 3-hour precipitation values at 15:00 and 18:00 to derive the 6-hour precipitation value at 18:00), the corresponding values are also incorporated into the final dataset.

\subsubsection{Neighboring Stations Check}

A common quality control method involves comparing the observations of a station with those of surrounding stations. WeatherReal-ISD merges neighboring stations that report consistent data, ensuring that the neighboring stations check does not use duplicate data for cross-validation, thereby enhancing the reliability. For any given station, stations within a 300 km distance and a 500m elevation difference are considered candidates. Candidates exhibiting less than one-third overlap in records with the station are filtered out first. The remaining candidates are categorized into four directional quadrants (northeast, northwest, southeast, southwest), with up to two nearest stations selected from each quadrant to help mitigate bias caused by large spatial variations over a local area. If at least three stations are selected in total, the neighboring stations check will be performed.

When comparing the observations of two stations, the same algorithm described in the Distributional Gap is employed. Then flags from all neighboring stations are aggregated. For each time step, it is still required that there be at least three records from neighboring stations. If two-thirds of the neighboring stations return the same result, the station under review will be flagged accordingly.

\begin{figure}[htbp]
    \centering
    \includegraphics[width=\linewidth]{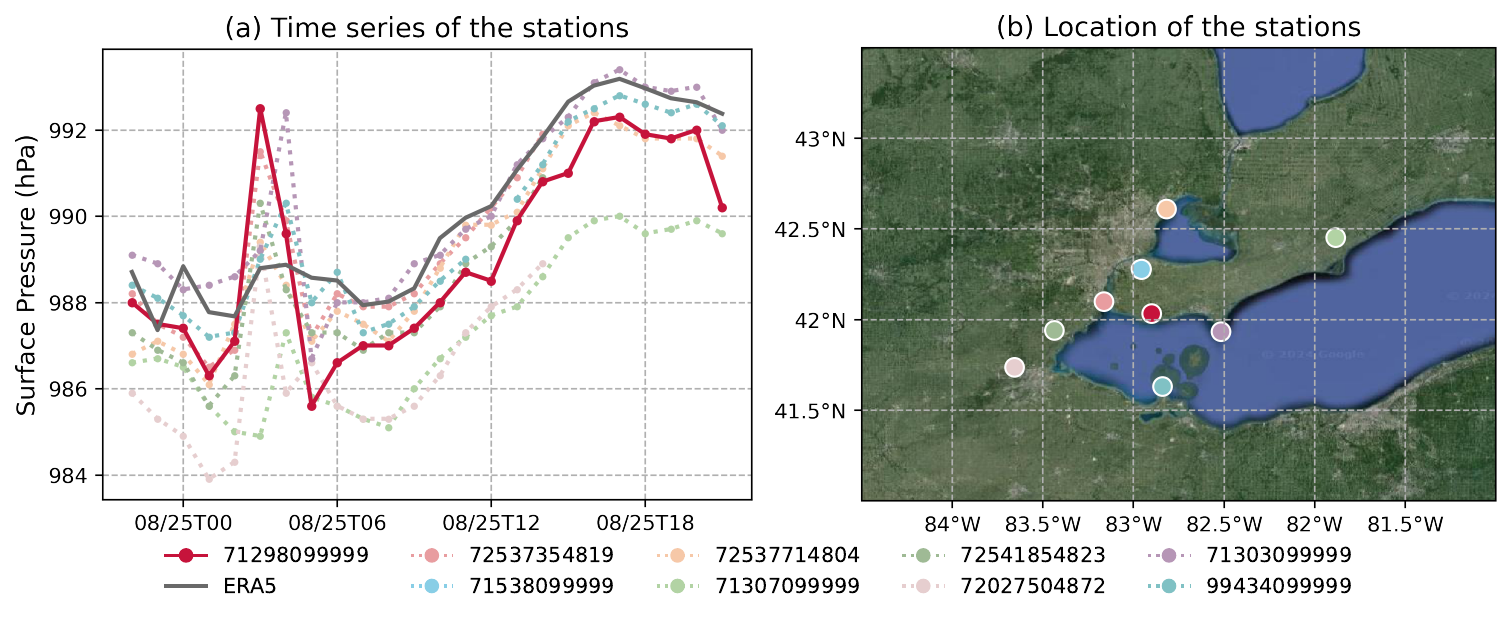}
    \caption{(a) Time series of surface pressure at station 71298099999 and its neighboring stations in 2023; (b) The spatial location of the station 71298099999 and its neighboring stations.}
    \label{fig:neighboring}
\end{figure}

One of the crucial roles of this method is to verify the flags of other algorithms. During the intra-station checks, there may be instances where valid records are incorrectly flagged. However, issues caused by equipment malfunctions or data transmission problems are unlikely to simultaneously occur at multiple stations. Therefore, comparing records from neighboring stations is a valuable supplementary method. The detailed cross-validation of all algorithms will be presented in Section~\ref{sec:integration}. Figure~\ref{fig:neighboring}a shows that station 71298099999 experienced a suspected pressure spike on Aug 25, 2023, which was not observed in the ERA5 data. However, since all eight neighboring stations in different directions (Figure~\ref{fig:neighboring}b) displayed the same pattern of change, the observation would not be flagged as erroneous by the spike check.

\subsubsection{Flag Refinement}

\begin{figure}[htbp]
    \centering
    \includegraphics[width=\linewidth]{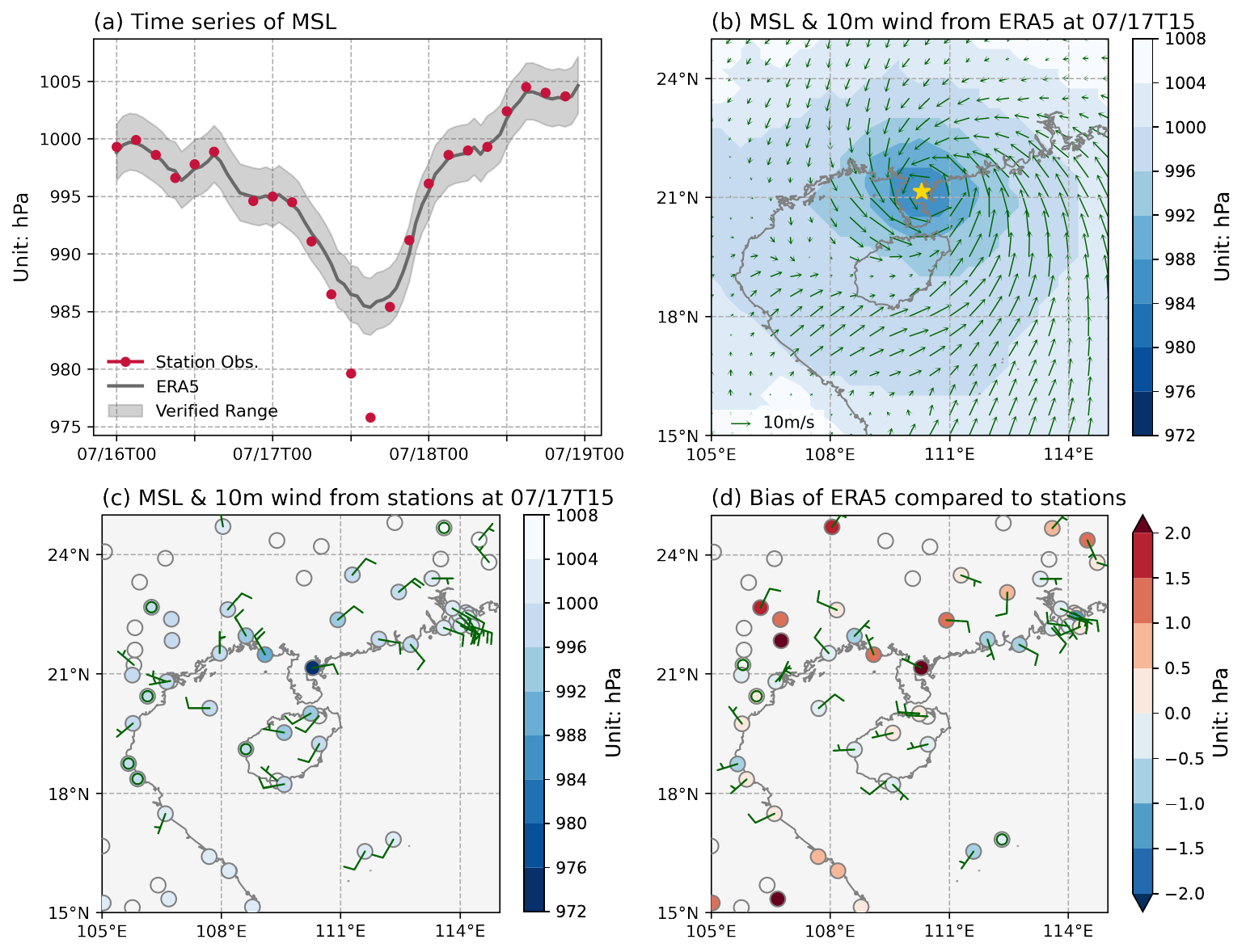}
    \caption{(a) Time series of mean sea-level pressure at station 59658099999 (Zhanjiang, China) during July 16-18, 2023; (b) The spatial distribution of mean sea-level pressure and wind vectors from ERA5 at 15:00 UTC on July 17, 2023, with a star indicating the location of the station; (c) Observations of mean sea-level pressure and wind barbs from nearby stations at the same timestamp; (d) The bias of mean sea-level pressure from ERA5 compared to station observations at the same timestamp.}
    \label{fig:low_pressure}
\end{figure}

\textbf{Low Pressure Minimum.} Manual inspection of the quality control effects on a large volume of observational data in 2023 indicates that the vast majority of flags correctly identified problematic data, with only a few exceptions. One typical scenario is that significantly low pressure records may be mistakenly identified as erroneous data when the station is located at the center of low pressure. Figure~\ref{fig:low_pressure}a presents the mean sea-level pressure time series at Zhanjiang Station from July 16 to 18. It can be observed that the minimum pressure during this period reached approximately 975 hPa at 15:00 UTC due to the impact of tropical storm Talim, with a difference of 10 hPa compared to ERA5, which exceeds the threshold for the Distributional Gap check. To address such issues at coastal stations, when the flagged pressure observation is a trough value within the day, while the observation values before and after both show a monotonic change, the observation flagged as errors by the Distributional Gap check or Spike check will be restored.

\textbf{Temperature Diurnal Cycle.} Temperature data from some stations often exhibit significant diurnal variations, making them more likely to be incorrectly flagged by Spike checks or Distributional Gap checks compared to other variables. Therefore, temperature data undergo an extra diurnal cycle check. A sine curve is fitted to the hourly temperature over a 24-hour period. The amplitude is determined by the maximum and minimum values within this period, and the phase is based on the sunrise time, with the minimum temperature restricted to between 3 hours before sunrise and 1 hour after sunrise. The sunrise time is calculated using the date and the station’s latitude and longitude. If the temperature matches the sine curve, the flags from other algorithms will be downgraded. This diurnal cycle check is only performed if the station is within 60°S-60°N and has at least two valid data points in each quartile of the day. Figure~\ref{fig:QC_step_check}e shows that at station 2861209999, the temperature dropped by 10.3°C between 20:00 and 23:00 local time on April 5, 2023, and then increased by 10.4°C between 8:00 and 11:00 on the next day. Although this situation is rare, it was not flagged as erroneous because the temperature changes aligned with the expected diurnal variation pattern.

\subsubsection{\label{sec:integration}Algorithm Integration}

The inspection results of each quality control algorithm are categorized into three types: normal, suspect, and erroneous. Only data points flagged as erroneous are removed. The criteria for each algorithm are detailed in Table~\ref{tab:qc_summary}. Flags from different algorithms undergo cross-validation to adjust their levels. For instance, the Spike Check initially flags identified spikes as suspect. These data are only flagged as erroneous if they are also flagged as suspect by either the Distributional Gap Check or the Neighboring Stations Check. Conversely, some erroneous flags may be downgraded to suspect by other algorithms, as demonstrated in the case in Figure~\ref{fig:QC_step_check}e and Figure~\ref{fig:neighboring}a.

\begin{table}[htbp]
\caption{Summary of quality control algorithms}
\label{tab:qc_summary}
\centering
\begin{tabular}{@{}>{\centering\arraybackslash}p{2cm}>{\centering\arraybackslash}p{.5cm}>{\centering\arraybackslash}p{.5cm}>{\centering\arraybackslash}p{1.8cm}>{\centering\arraybackslash}p{.5cm}>{\centering\arraybackslash}p{1.8cm}>{\centering\arraybackslash}p{.5cm}>{\raggedright\arraybackslash}p{5.5cm}@{}}
\toprule
Algorithm & t & td & sp \& msl & c & ws \& wd & Ra & Flag Level \\
\midrule
Known records & \checkmark & \checkmark & \checkmark & \checkmark & \checkmark & \checkmark & All outliers are flagged as erroneous. \\
\midrule
Distributional gap & \checkmark & \checkmark & \checkmark & & & & Records falling outside the 0.01 probability density range are suspect; Records falling outside the identified gap are erroneous.\\
\midrule
Spike & \checkmark & \checkmark & \checkmark & & & & Flag as erroneous only if all records are flagged as at least suspect by either the distributional gap algorithm or neighboring stations.\\
\midrule
Persistence & \checkmark & \checkmark & \checkmark & \checkmark & \checkmark & \checkmark & Flag as erroneous only when it is longer than 48h or contains records flagged by other algorithms. \\
\midrule
Cluster deviation & \checkmark & \checkmark & \checkmark & & & & Flags are only used for calculation of distributional gap. \\
\midrule
Super-saturation & \checkmark & \checkmark & & & & & All detected records are flagged as erroneous. \\
\midrule
Wind consistency & & & & & \checkmark & &  All detected records are flagged as erroneous. \\
\midrule
Precipitation consistency & & & & & & \checkmark &  All detected records are flagged as erroneous. \\
\midrule
Neighboring stations & \checkmark & \checkmark & \checkmark & & & & Records falling outside the 0.01 probability density range are suspect; Records falling outside the identified gap are erroneous. \\
\midrule
Low pressure minimum & & & \checkmark & & & & All detected erroneous flags are downgraded to suspect. \\
\midrule
Temperature diurnal cycle & \checkmark & & & & & & Only used for downgrading flags from other algorithms. \\
\bottomrule
\end{tabular}
\end{table}

\begin{figure}[htbp]
    \centering
    \includegraphics[width=\linewidth]{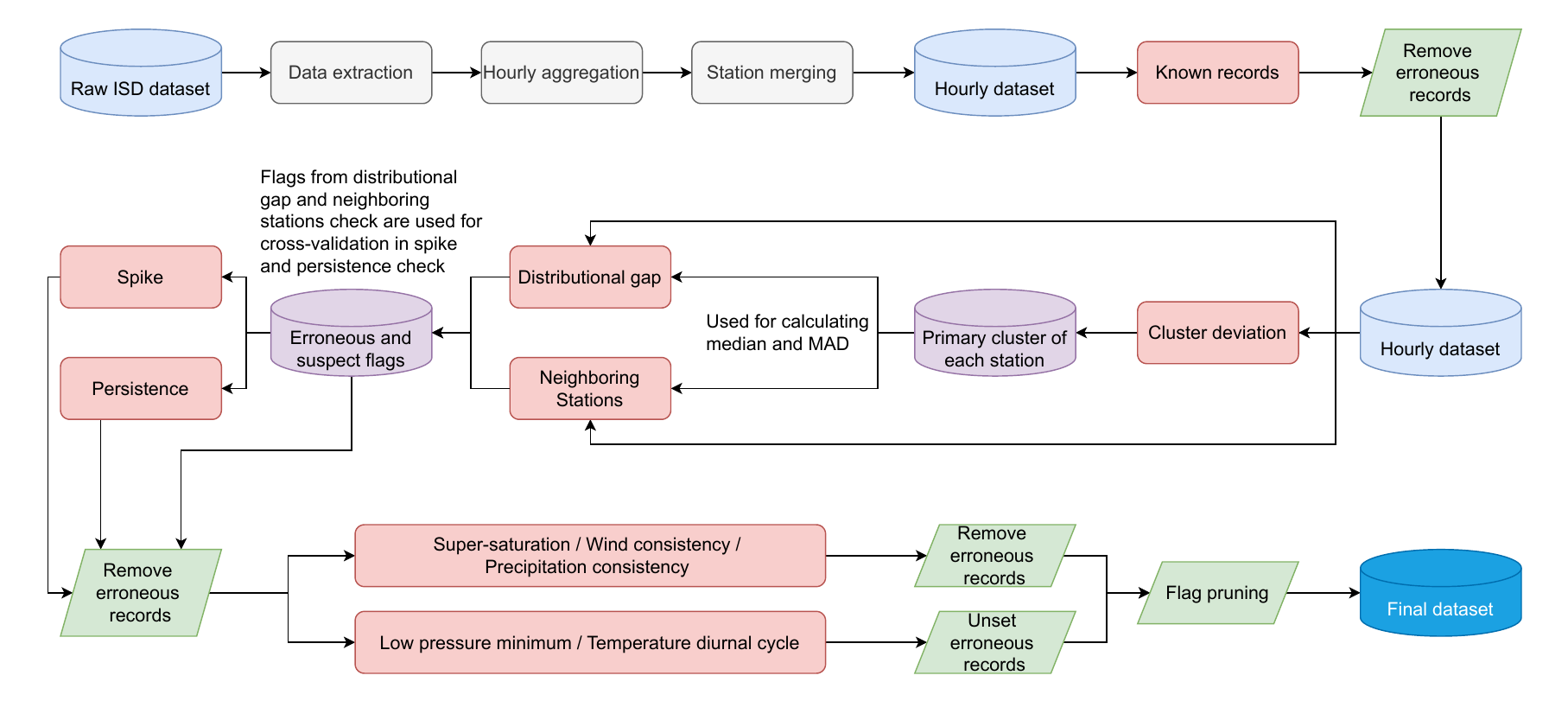}
    \caption{Flow diagram of the quality control procedure. The scope differs across algorithms and certain steps may be omitted for different variables.}
    \label{fig:qc_diagram}
\end{figure}

In addition to cross-validation, the flags from different algorithms are combined and pruned before the ultimate dataset is generated. For any segment flagged as suspect, if the adjacent records on both sides are flagged as erroneous, all the suspect flags within the segment are upgraded to erroneous. When within a one-month sliding window, more than half of the data points are flagged as suspect or erroneous, all suspect flags in this window are upgraded to erroneous. For instance, temperature data at some stations may exhibit frequent and random abnormal fluctuations over a certain period, making it difficult to be accurately flagged by the Spike Check, thus suspect data are also excluded. In addition, if more than half of the data points at a station are flagged as erroneous, the station is removed. A typical situation is when the data recorded at a station remain the same value for most of the time, resulting in low reliability for all remaining records. The complete procedure is shown as a flow diagram in Figure~\ref{fig:qc_diagram}.

\section{Dataset Analysis}

\subsection{Station density and reporting frequency}

In contrast to reanalysis data, in-situ observation data exhibits varying distribution densities and reporting frequencies, influenced by factors such as terrain, population density and economic resources. 

\begin{figure}[htbp]
    \centering
    \includegraphics[width=\linewidth]{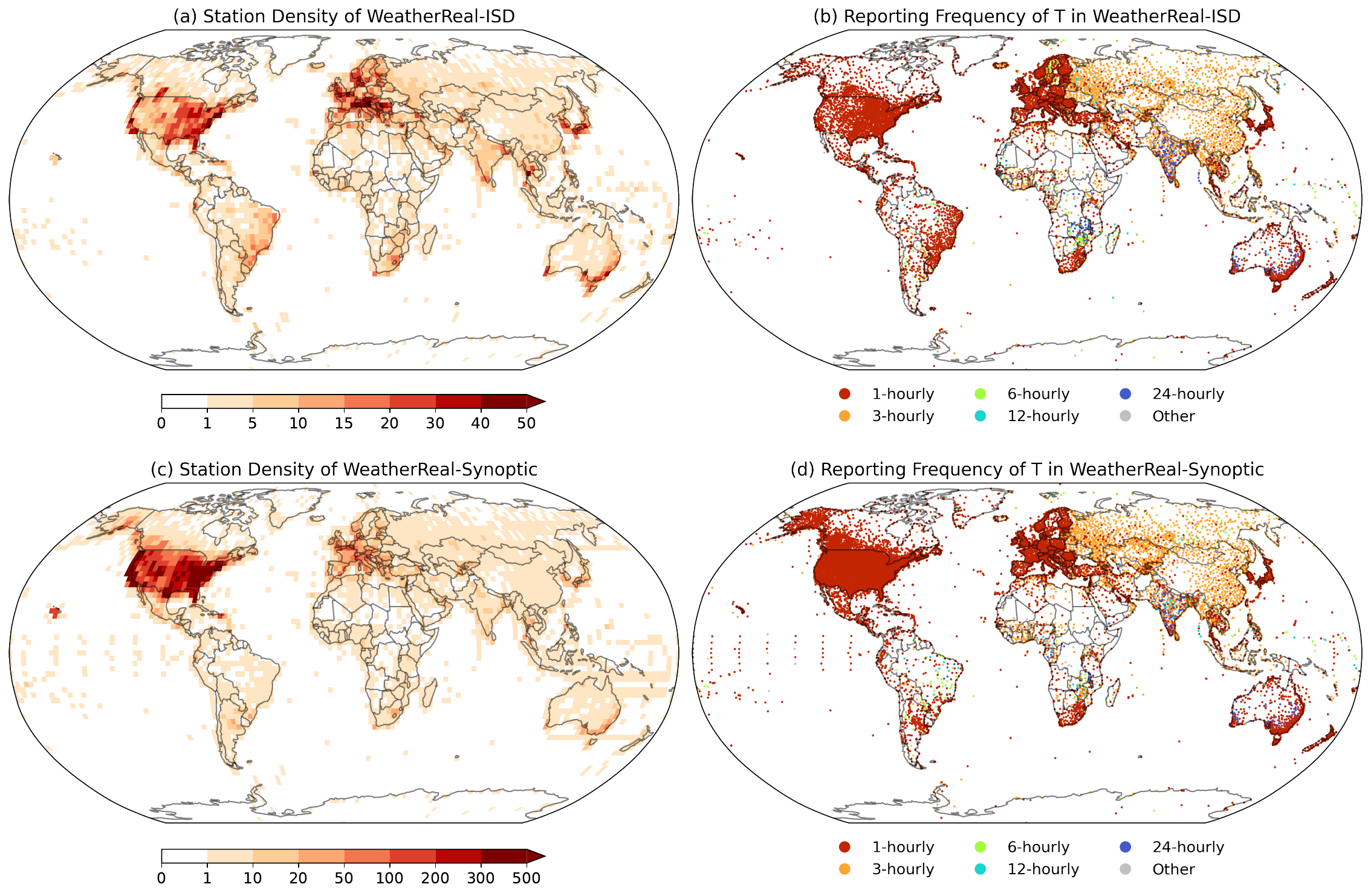}
    \caption{Station density and 2m temperature reporting frequency for WeatherReal-ISD (top row) and WeatherReal-Synoptic (bottom row) in 2023. Station density (left column) is shown as the number of stations within 2.5°×2.5° grids. Note that (a) and (c) use different scales for the color bars. “Other” in the reporting frequency (right column) indicates sporadic records.}
    \label{fig:station_density_and_frequency}
\end{figure}

Figure~\ref{fig:station_density_and_frequency}a shows the global distribution of WeatherReal-ISD station density in 2023. Densely located stations are primarily concentrated in the United States, particularly in the eastern region, Western to Southern Europe, East Asia, especially Japan, and the southeastern coastal areas of Australia. Conversely, regions such as South America, central Africa, and the Tibetan Plateau are severely lacking in the distribution of stations. Figure~\ref{fig:station_density_and_frequency}b shows the reporting frequency of 2m temperature, the parameter with the highest availability, of WeatherReal-ISD in 2023. The reporting frequency for each station is determined by calculating the median interval between two consecutive records within 2023. Given that the reporting frequency for the vast majority of stations is 1, 3, 6, 12, or 24 hours, the few stations with uncommon frequencies are reclassified to the lower frequency category. It can be seen that regions with higher station density often have higher reporting frequencies. Stations in the United States, Europe, and Japan predominantly report hourly. In contrast, stations in China, Central Asia, and Russia generally report every three hours. Regions such as India, South America, and Africa exhibit even lower reporting frequencies.

On the other hand, WeatherReal-Synoptic is distinguished by a significantly higher density of stations in the United States and Europe (Figure~\ref{fig:station_density_and_frequency}c). This is attributable to Synoptic’s integration of numerous observation networks, such as Citizen Weather Observer Program (CWOP), a private-public partnership aimed at collecting weather data contributed by citizens. Although the quality of these observations can vary considerably, they offer highly dense real-time data. As shown in Figure~\ref{fig:station_density_and_frequency}d, compared to ISD, Synoptic has a significantly higher number of stations reporting hourly data in the western United States.

Table~\ref{tab:frequency} lists the number of stations in WeatherReal with different reporting frequencies. Precipitation data is classified uniquely: As stations may record accumulated precipitation in multiple intervals, they are classified based on the shortest interval. In WeatherReal-ISD, the number of stations recording temperature, dewpoint, and wind exceeds 12,000, with over 8,000 stations recording at hourly intervals. Conversely, other parameters are fewer in number and record at lower frequencies. For WeatherReal-Synoptic, the integration of additional observation networks has resulted in over 30,000 stations for measuring temperature, wind, and precipitation. Specifically, there are more than 40,000 stations that record temperature on an hourly basis. The number of stations for dewpoint, surface pressure, and mean sea-level pressure is lower, possibly due to the higher measurement requirements for these parameters. Moreover, the dewpoint and mean sea-level pressure data here are derived from other parameters (e.g., dewpoint is calculated from temperature and relative humidity only when both of them are recorded at a station). WeatherReal-Synoptic also includes nearly 20,000 daily precipitation stations, most of which are part of observation networks like CWOP and the Cooperative Observer Network (COOP). Also note that the numbers for WeatherReal-Synoptic presented here are based on the subset of data we collected in 2023, and may be lower than the total provided by the Synoptic service. The table does not include cloud from Synoptic. Although this variable is also available, it may not have an advantage in the number of stations compared to ISD, possibly due to fewer observation networks equipped with instruments to observe cloud cover.

\begin{table}[htbp]
\caption{Number of stations in with different reporting frequencies for each variable in 2023}
\label{tab:frequency}
\centering
\begin{threeparttable}
\begin{tabular}{@{}lllllllll@{}}
\toprule
Dataset & Variable & Total & 1-hourly & 3-hourly & 6-hourly & 12-hourly & 24-hourly & Other \\
\midrule
WeatherReal-ISD & t & 12901 & 8655 & 2913 & 419 & 125 & 477 & 312 \\
& td & 12387 & 8187 & 2908 & 403 & 120 & 453 & 316 \\
& sp & 9919 & 5546 & 3589 & 307 & 115 & 170 & 192 \\
& msl & 9583 & 5186 & 3333 & 319 & 129 & 269 & 347 \\
& ws \& wd & 12916 & 8510 & 2921 & 405 & 118 & 479 & 483 \\
& c & 8981 & 4985 & 2584 & 355 & 148 & 510 & 399 \\
& ra & 9600 & 3871 & 869 & 3023 & 1245 & 592 & N/A \\
\midrule
WeatherReal-Synoptic & t & 53325 & 42038 & 6291 & 518 & 124 & 3699 & 655 \\
& td & 29671 & 23913 & 968 & 899 & 750 & 388 & 2752 \\
& sp & 17037 & 11966 & 1092 & 770 & 462 & 281 & 2466 \\
& msl & 19578 & 13838 & 1302 & 875 & 497 & 364 & 2701 \\
& ws \& wd & 44393 & 37556 & 5448 & 446 & 130 & 344 & 469 \\
& ra & 34214 & 10639 & 335 & 4155 & N/A & 19085 & N/A \\
\bottomrule
\end{tabular}
\end{threeparttable}
\end{table}

\subsection{Overall data validation}

\begin{figure}[htbp]
    \centering
    \includegraphics[width=\linewidth]{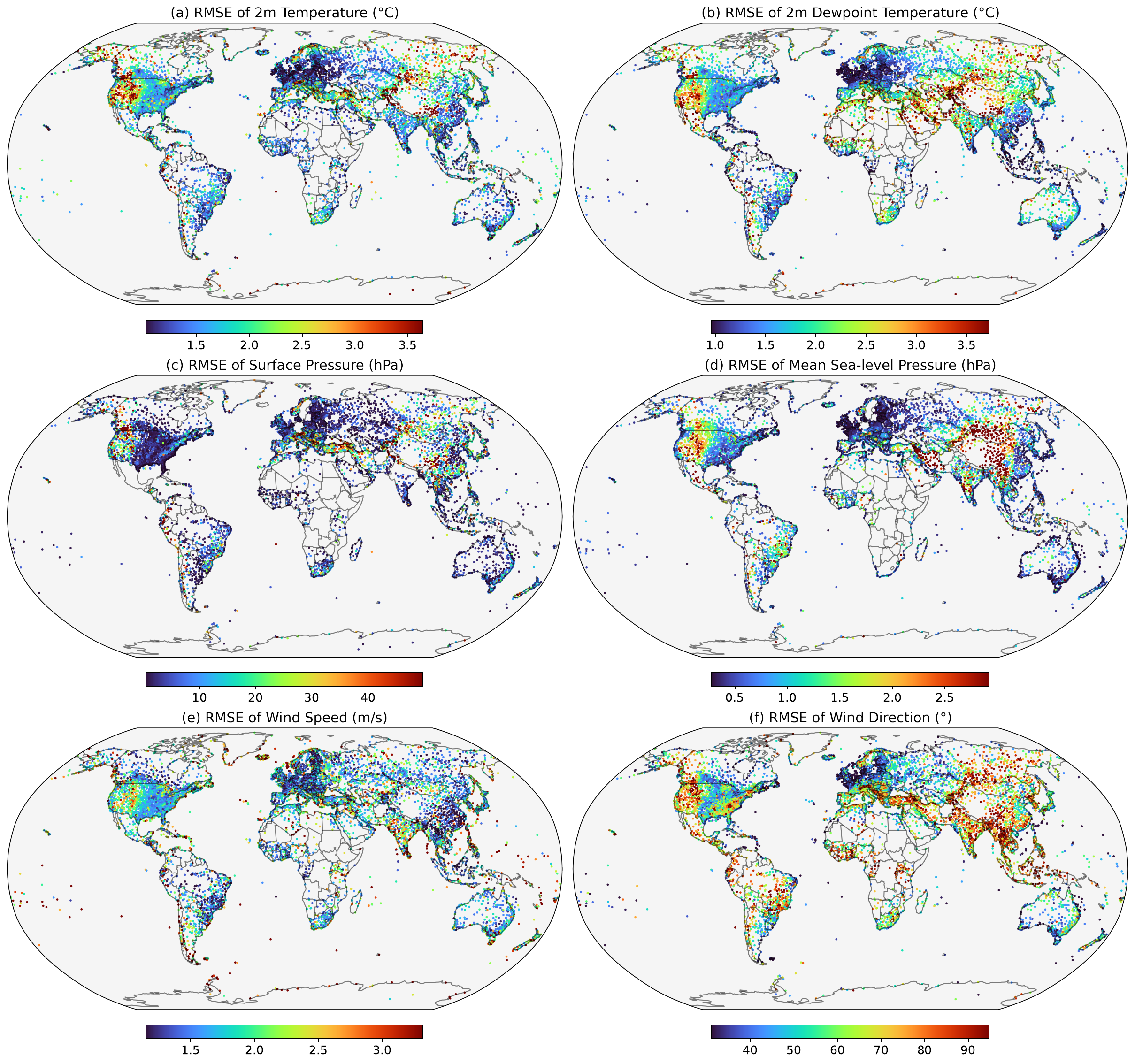}
    \caption{RMSE between WeatherReal-ISD and ERA5 for the entire year of 2023.}
    \label{fig:rmse}
\end{figure}

We begin by showing some basic sanity checks for the WeatherReal observation datasets compared to a widely used source of ground truth, ERA5. The grid data from ERA5 is bilinearly interpolated to match the latitude and longitude coordinates of the stations. Figure~\ref{fig:rmse} presents the root mean squared error (RMSE) between WeatherReal-ISD and ERA5 reanalysis data for various variables, including 2m temperature, 2m dewpoint temperature, surface pressure, mean sea-level pressure, 10m wind speed, and wind direction. The primary source of the discrepancy is likely the influence of terrain and resolution. ERA5 provides grid analysis within cells at a 0.25° resolution, which tends to smooth out local extremes.

For 2m temperature and dewpoint temperature (Figure~\ref{fig:rmse}a-b), the RMSE in most regions is within 2 degrees. Stations with higher RMSE are concentrated in high-altitude regions such as western North America and from Central Asia to western China. The ERA5 reanalysis may struggle to capture the fine-scale variations in temperature fields in these regions. For surface and mean sea-level pressure (Figure~\ref{fig:rmse}c-d), the RMSE is also predominantly higher in regions with complex topography. It is noted that the RMSE of surface pressure is significantly larger than that of mean sea-level pressure, mainly due to the considerable discrepancies between the topography represented in ERA5 and the actual elevation at the station locations. Applying an elevation correction would reduce the RMSE by an order of magnitude (not shown). The RMSE for wind speed and direction (Figure~\ref{fig:rmse}e-f) shows more spatial variability compared to the former parameters, with wind direction errors being particularly pronounced. Unlike other variables, wind direction differences are significant in coastal areas such as the eastern United States, southern Europe, and South Asia. Notably, in regions with high station density, such as the eastern United States, western Europe, East Asia, and coastal areas of Australia, the differences between station observations and ERA5 are often smaller. This may indicate that in areas with dense station observations, the data assimilation in ERA5 is more effective.

\begin{figure}[htbp]
    \centering
    \includegraphics[width=\linewidth]{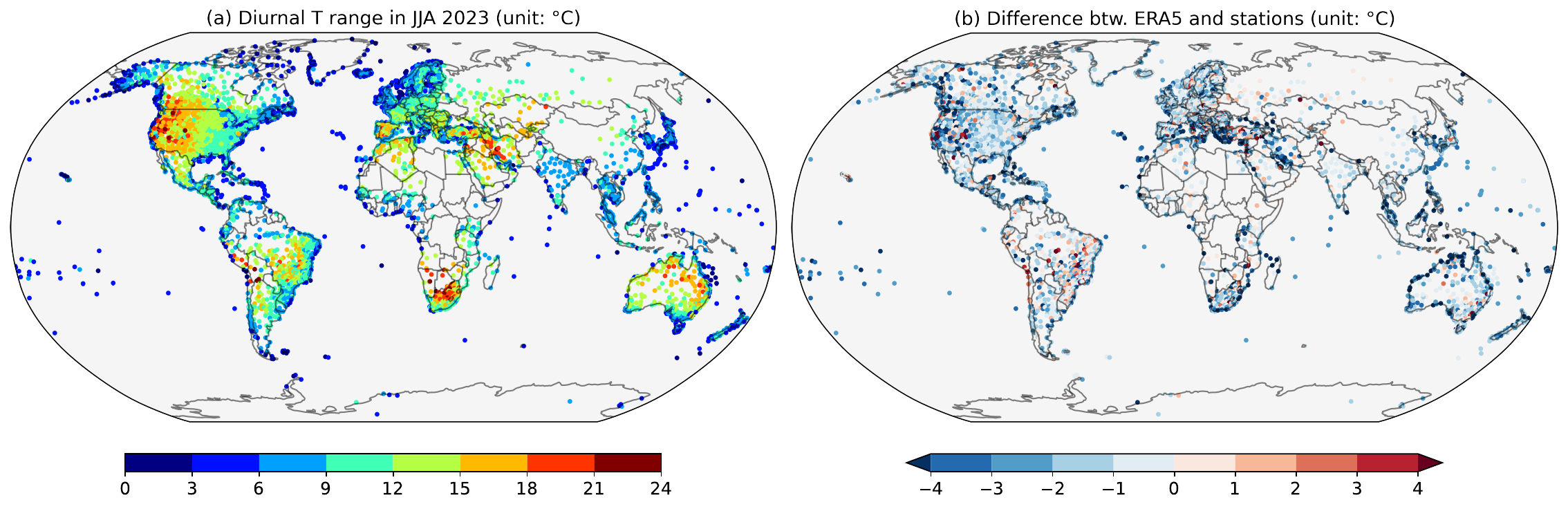}
    \caption{(a) DTR of station observation from WeatherReal-ISD in JJA, 2023; (b) The difference of DTR between ERA5 and WeatherReal-ISD in JJA, 2023.}
    \label{fig:diurnal}
\end{figure}

For 2m temperature, the diurnal temperature range (DTR) is also analyzed. Figure~\ref{fig:diurnal}a shows the median DTR at different stations during summer (JJA) in 2023. Only hourly and 3-hourly stations are plotted to ensure more accurate daily maximum and minimum temperatures. It can be seen that high-altitude regions, such as the western United States, southern Africa, and arid regions, such as the Middle East and inland Australia, have higher DTR, while coastal areas tend to have lower DTR. A notably lower DTR in ERA5 is revealed in Figure~\ref{fig:diurnal}b. This discrepancy is evident across 80\% of global stations, with particularly significant differences observed in regions such as western North America and southern Europe. It suggests that ERA5 may under-represent daily temperature variability, potentially due to its spatial resolution or modeling approach. This highlights the importance of using hyper-local station data for understanding and modeling local weather dynamics.

To further investigate the sources of error in ERA5, we utilized WeatherReal-Synoptic to evaluate the average bias between station observations and ERA5 data. The analysis concentrated on the contiguous United States (CONUS) due to its high density of Synoptic stations.

\begin{figure}[htbp]
    \centering
    \includegraphics[width=\linewidth]{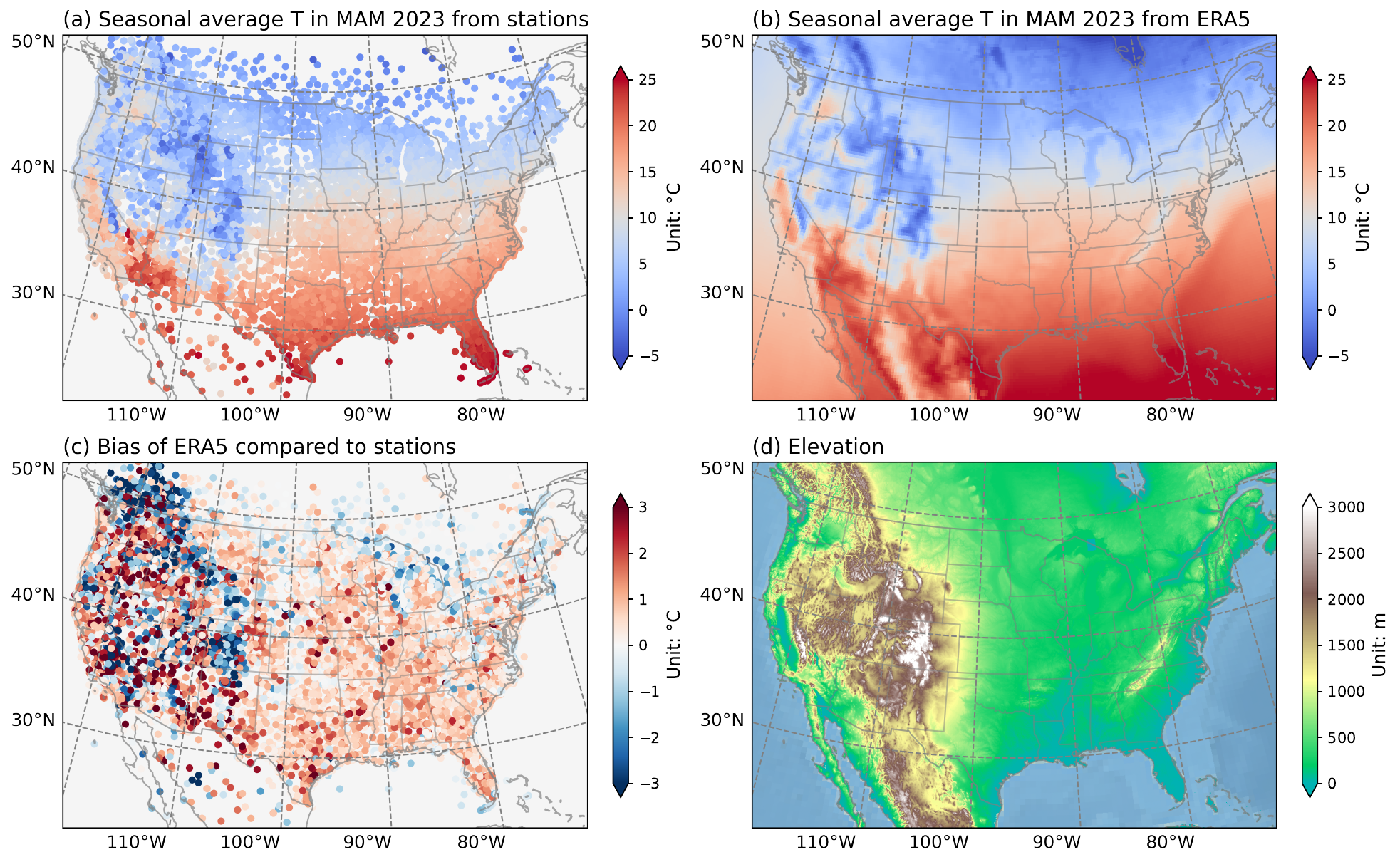}
    \caption{Seasonal average 2m temperature in MAM, 2023 of (a) station observations from WeatherReal-Synoptic (b) ERA5 and (c) their difference; (d) The elevation from Shuttle Radar Topography Mission (SRTM)~\citep{farr2007shuttle}.}
    \label{fig:synoptic_MAM_T}
\end{figure}

Figure~\ref{fig:synoptic_MAM_T}a shows the spatial distribution of the seasonal average temperature from WeatherReal-Synoptic for the spring of 2023 (MAM). Due to a more extensive observation network, WeatherReal-Synoptic covers almost the entire CONUS with more than 21000 hourly stations in 2023, reflecting the large-scale temperature distribution, which are consistent with ERA5 (Figure~\ref{fig:synoptic_MAM_T}b) and local terrain-influenced temperature variations. The differences between ERA5 and station observations  (Figure~\ref{fig:synoptic_MAM_T}c) reveal clusters of significant warm and cold biases in the western United States, closely related to elevation (Figure~\ref{fig:synoptic_MAM_T}d). In regions with higher elevation, such as along the Rocky Mountains and the Sierra Nevada, station-observed temperatures are often higher than those in ERA5, while temperatures on the mountain sides are lower. In the central and eastern CONUS, biases are smaller, but ERA5 generally shows a positive bias at most station locations. Conversely, along the northeastern shores of the Great Lakes, ERA5 underestimates the average spring temperature.

\begin{figure}[htbp]
    \centering
    \includegraphics[width=\linewidth]{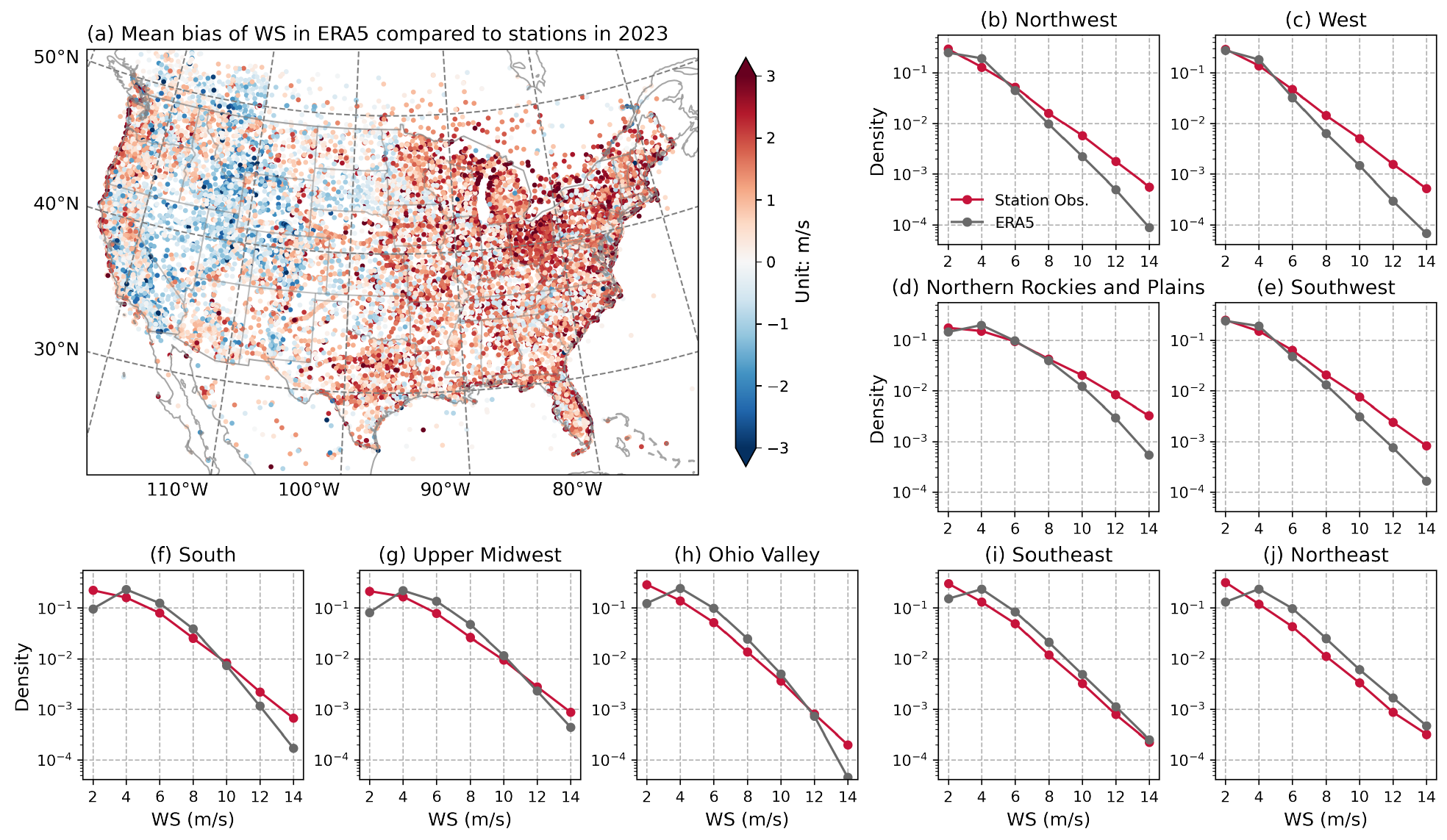}
    \caption{(a) The mean bias of wind speed in ERA5 compared to station observations from WeatherReal-Synoptic in 2023; (b-j) The density distribution of all wind speed values for WeatherReal-Synoptic and ERA5 in nine different climate regions in 2023, with the Y-axis displayed on a logarithmic scale. The climate regions are divided by states as described in \cite{karl1984regional}.}
    \label{fig:synoptic_WS}
\end{figure}

WeatherReal-Synoptic also collected hourly wind speed observations from over 18,000 stations across the CONUS. The average wind speed bias between ERA5 and station observations in this year (Figure~\ref{fig:synoptic_WS}a) reveals that ERA5 generally overestimates wind speeds in much of the eastern CONUS, particularly in the northeastern states, where the bias frequently exceeds 2 m/s. In contrast, the western CONUS exhibits a more complex pattern, with most regions where ERA5 underestimates wind speeds compared to station observations. To reveal the main sources of the bias, Figure~\ref{fig:synoptic_WS}b-j show the hourly wind speed value frequency distributions for different regions of the CONUS. For the western CONUS (Figures b-e), ERA5 overestimates the frequency of wind speeds below 4 m/s and underestimates the frequency above 4 m/s. This explains the negative wind speed bias in most parts of the western CONUS. In the central regions of the CONUS (Figure~\ref{fig:synoptic_WS}f-h, ERA5 continues to underestimate the frequency of strong winds exceeding 12 m/s.This pattern of overestimating weak winds and underestimating strong winds has been noted in previous studies~\citep{schewe2019SOTA, molina2021comparison}. Consequently, it may affect the evaluation of extreme weather forecasts based on ERA5. For the eastern coastal region, the wind speed density distribution in ERA5 aligns more closely with the observations.

\subsection{Case studies}

A further investigation targeting extreme events, where hyper-local in-situ observations are most important, can help emphasize the WeatherReal datasets' representation of user-impacting weather phenomena. Furthermore, this analysis will also highlight the efficacy of the dataset quality control algorithms discussed in Section~\ref{section:qc}. Heatwave, heavy rainfall and tropical cyclone events in 2023 are selected to analyze the dataset’s utility using the WeatherReal-ISD and user reports.

\subsubsection{Heatwave, Southeast Europe, July 2023}

Much of southeastern Europe, and parts of western and central Europe, saw their warmest year on record in 2023. In July, southeastern Europe experienced a prolonged heatwave, with 41\% of southern Europe, experiencing “strong”, “very strong” or “extreme” heat stress on 23 July. Greece, in particular, experienced its largest ever recorded wildfire~\citep{Copernicus2023}.

\begin{figure}[htbp]
    \centering
    \includegraphics[width=\linewidth]{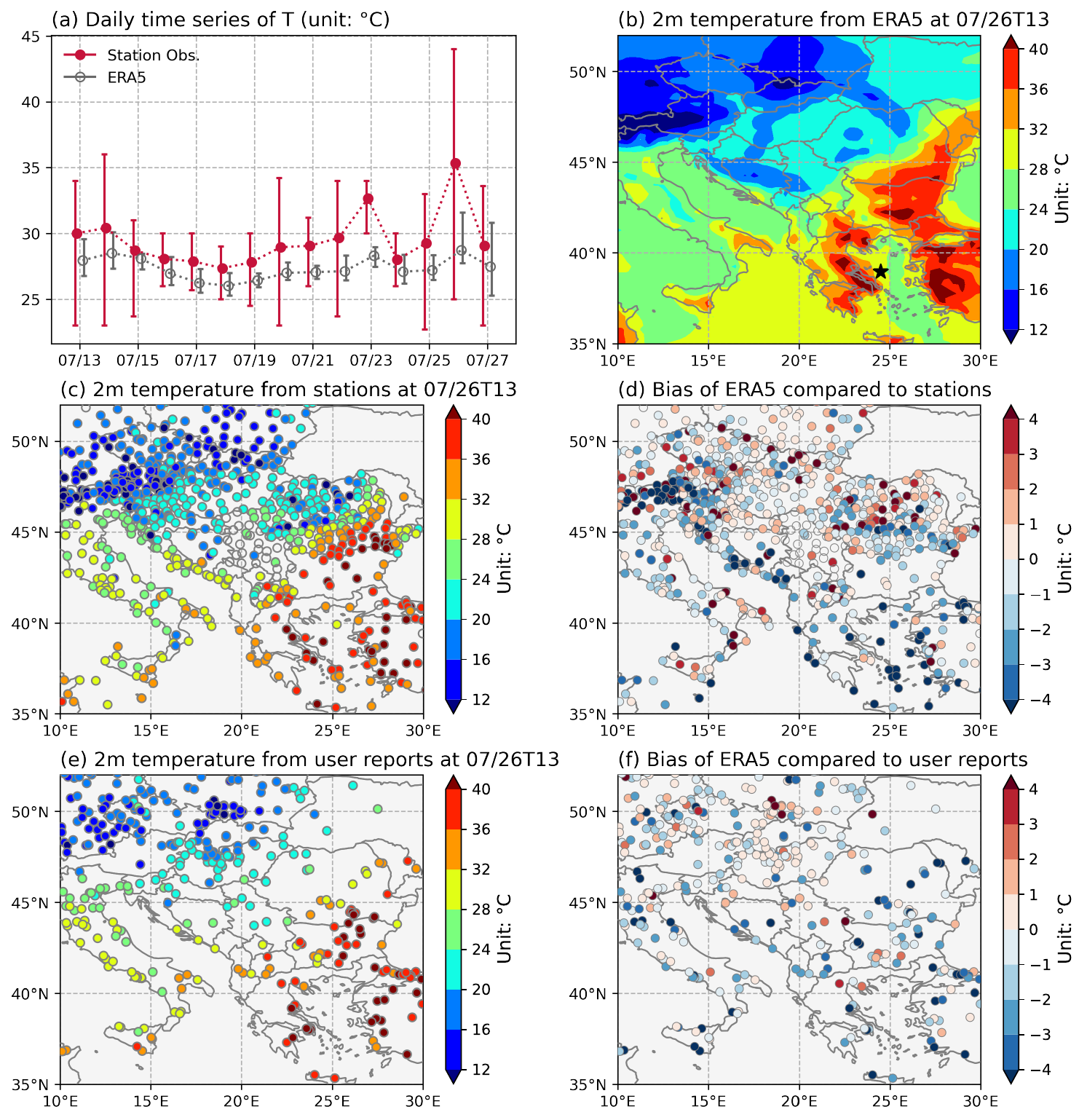}
    \caption{(a) The daily 2m temperature at station 16684099999 in Greece from July 13 to 27, where the upper (lower) end of the bar represents the daily maximum (minimum), and the point in the middle represents the daily average. For better visual representation, the bars representing station observation and ERA5 are slightly shifted to the left and right, respectively. (b) The spatial distribution of temperature from ERA5 at 13:00 UTC on July 26, 2023, with a star indicating the location of the station; (c) The spatial distribution of temperature from stations at the same timestamp; (d) The temperature difference between station observation and ERA5 at the same timestamp. (e-f) Similar to (c-d) but for user reports.}
    \label{fig:case_heatwave}
\end{figure}

Figure~\ref{fig:case_heatwave}a illustrates the daily temperature variations at Skyros in Greece (16684099999) from July 13 to 27. On July 26, the station recorded a maximum temperature of 44.0°C, which was 11.0°C higher than the previous day. In contrast, the ERA5 data for the same location on July 26 showed a maximum temperature of only 31.6°C, underestimating the station’s observed temperature by 12.4°C. Additionally, the figure indicates that ERA5 consistently underestimated the diurnal temperature range at this station throughout the period. The diurnal range reported by ERA5 remained mostly within 3°C, whereas the station observation often showed a range exceeding 10°C.

Figure~\ref{fig:case_heatwave}b-d respectively show the spatial distribution of temperature from ERA5, station observations, and their differences at 13:00 UTC (16:00 local time in Greece) on July 26 in Europe. It can be observed that while both datasets display similar spatial patterns, ERA5 significantly underestimates the high temperature in the southern coastal regions of Europe. The user reports collected by MSN Weather are further utilized to validate the comparison results here (Figure~\ref{fig:case_heatwave}e-f). Since the temperature reports are provided as ranges (typically 2-4°C) rather than exact values, the average of these ranges is used. It can be observed that, compared to the user reports, ERA5 still shows a cold bias across much of southeastern Europe.

The discrepancy can be attributed to the region’s complex terrain, characterized by numerous islands and its proximity to the sea. The data from WeatherReal are more effective in capturing the sharp temperature swings due to the rapid heating and cooling of the land compared to the surrounding sea.

\subsubsection{Heavy rainfall, Southeast United States, August 2023}

\begin{figure}[htbp]
    \centering
    \includegraphics[width=\linewidth]{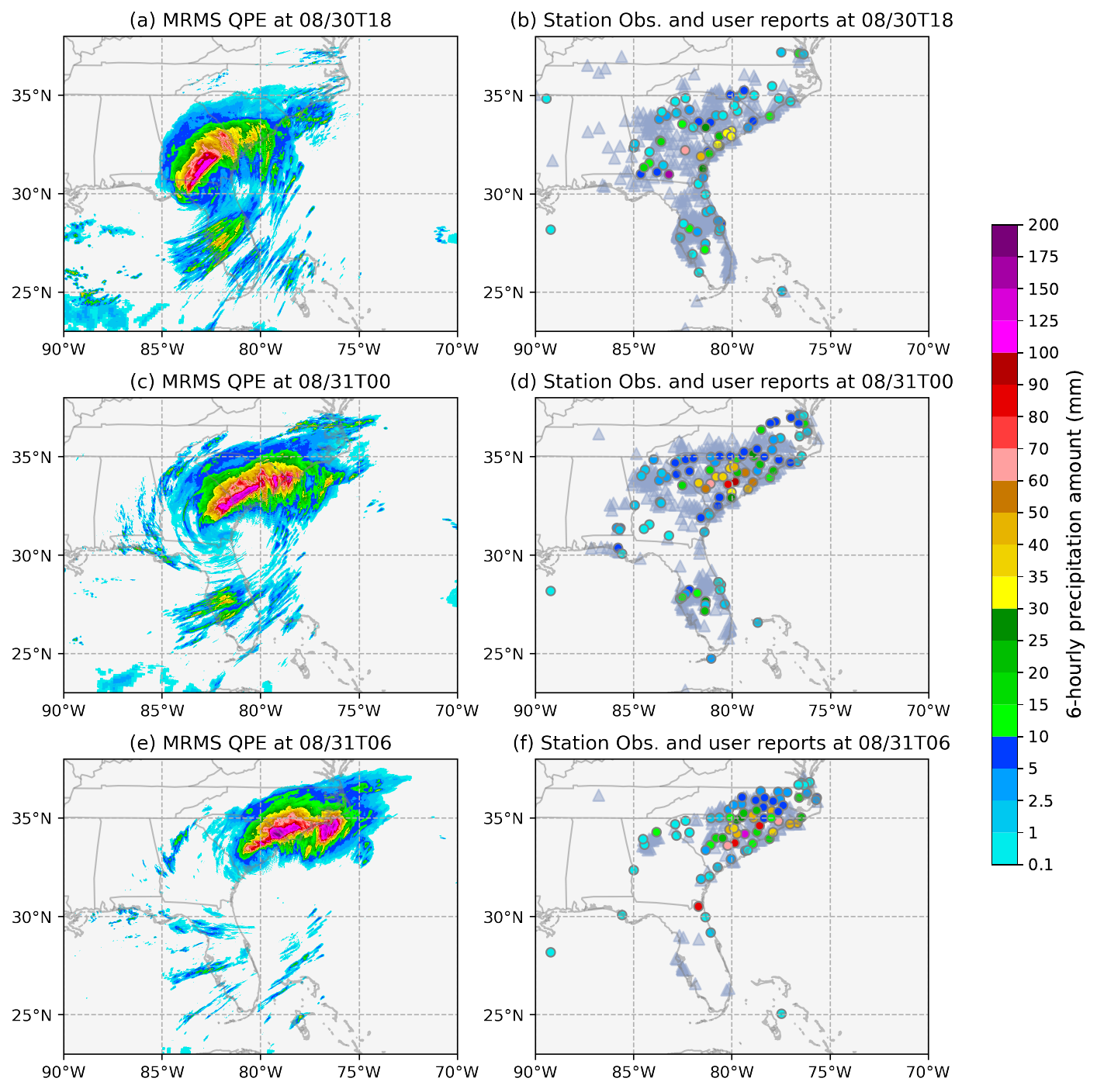}
    \caption{The spatial distribution of 6-hour accumulated precipitation from August 30-31, 2023. The left column presents MRMS QPE Pass 2, while the right column includes station observations and user reports. In the right column, circles denote station observations, and triangles denote user reports. User reports are binary data, with only those indicating precipitation in the last 6 hours being marked.}
    \label{fig:case_rainfall}
\end{figure}

Hurricane Idalia made landfall in Florida on August 30, 2023, crossed Georgia, and entered the Atlantic from the Carolinas on August 31, 2023~\citep{NWS2023}. The storm brought heavy rainfall to the region, lasting for an entire day. Figure~\ref{fig:case_rainfall} illustrates the 6-hour cumulative precipitation for three different time periods on August 30-31, 2023. Due to significant discrepancies between ERA5 precipitation data and the actual observations (not shown), the QPE (Quantitative Precipitation Estimation) pass 2 data from MRMS (Multi-Radar/Multi-Sensor)~\citep{MRMS2016} is used for comparison. As shown in Figure~\ref{fig:case_rainfall}, MRMS and station observation are largely consistent, both effectively illustrating the movement of the rain belt under the influence of Idalia.

In the right column of the figure, only stations with complete observations and reporting at least 0.1 mm of precipitation are plotted. The maximum 6-hour precipitation was recorded at station 74781013857 in Georgia, with 166.9 mm from 12:00 to 18:00 on August 30. Although precipitation records are less complete compared to other variables, the precipitation amounts observed at the stations, being direct measurements, should be more accurate than radar estimates, and certainly more so than ERA5 data.

The user reports collected by MSN Weather are also shown in the right column of Figure~\ref{fig:case_rainfall}, with triangles marking the locations where users reported precipitation in the past 6 hours. Although user reports do not include specific precipitation amounts, their distribution effectively shows the changes in major precipitation locations over the 24-hour period. Within the area depicted on the map (20° by 20° range), over 6,000 user reports were collected, surpassing the number of 1-hour precipitation records from stations. Additionally, user reports are not confined to station locations, making their distribution more random and reflective of the spatial distribution of user density.

\subsubsection{Typhoon Talim, Southeast China, July 2023}

On July 17, 2023, Talim, a tropical storm which formed near the Philippines, was upgraded to a typhoon. Around 14:20 UTC on the same day, the center of Talim made landfall along the coast of Zhanjiang, China~\citep{he2023observational}.
Figure~\ref{fig:low_pressure}b illustrates the spatial distribution of mean sea-level pressure and 10m wind in the southeastern coastal region of China at the time of the typhoon’s landfall based on ERA5. As previously mentioned, ERA5 significantly underestimated the low pressure intensity at the typhoon’s center.

Meanwhile, the mean sea-level pressure and 10m wind from station observations (Figure~\ref{fig:low_pressure}c) also illustrating the spatial distribution characteristics of the low-pressure cyclone. The mean sea-level pressure at the central station drops to 975.8 hPa. Stations with higher wind speeds are concentrated to the northeast of the low-pressure center, which is the right front quadrant of the typhoon’s movement. In this region, the typhoon’s rotational wind aligns with its movement direction, resulting in the highest wind speeds.

Apart from the central intensity, the characteristics depicted by station observations are basically consistent with those shown by ERA5. However, ERA5 shows higher pressure values inland and lower pressure values over the southeastern sea surface (Figure ~\ref{fig:low_pressure}d). Furthermore, discrepancies in wind direction suggest that the cyclonic intensity in ERA5 is slightly stronger than indicated by the actual observations. This discrepancy can also be attributed to the fact that modeled winds from ERA5 are less influenced by local effects compared to station observations. Consequently, station observation tends to exhibit more variability and noise, reflecting local influences.

\section{Example Evaluations}

\label{section:evaluation}

The first evaluation here shows how WeatherReal can be used to compare multiple types of forecasts, with a specific emphasis on evaluating the new Aurora model~\citep{bodnar_aurora_2024}. Included in this example are:
\begin{enumerate}
    \item \textbf{Aurora-9km}: The high-resolution fine-tuned version of the Aurora model from \cite{bodnar_aurora_2024}. The data-driven model is initialized with operational analysis from the IFS, and predicts at 6-hourly time intervals. Due to limitations in the inference run for this dataset as of this writing, only initialization times of Jan 2, 2023 through Dec 19, 2023, all at 00 UTC, are evaluated. For direct comparison in this example, other models are also limited to the same forecast issue times.
    \item \textbf{ECMWF}: The operational run of the ECMWF IFS model.
    \item \textbf{MS-Point}: Forecasts from the operational consumer weather forecasts produced by Microsoft and embedded into Microsoft weather services such as MSN Weather. The underlying model is an NWP post-processor that is based on station observations and frequently fine-tuned on new data. Realistically, MS-Point is at a slight disadvantage in this evaluation because it relies on data from ECMWF, which is only streamed with approximately 8-hour latency. Hence its predictions are based on an earlier (12 UTC) cycle of the IFS. For a fair comparison of MS-Point's post-processing capability, it should be compared to the same IFS and Aurora cycle.
    \item \textbf{GFS}: The operational Global Forecast System from the National Oceanic and Atmospheric Administration. This model is at coarser resolution (available at 0.25$^\circ$).
\end{enumerate}

\subsection{WeatherReal-ISD}
In this section, the forecasts are evaluated against WeatherReal-ISD for 2023.

\begin{figure}
    \centering
    \includegraphics[width=0.33\textwidth]{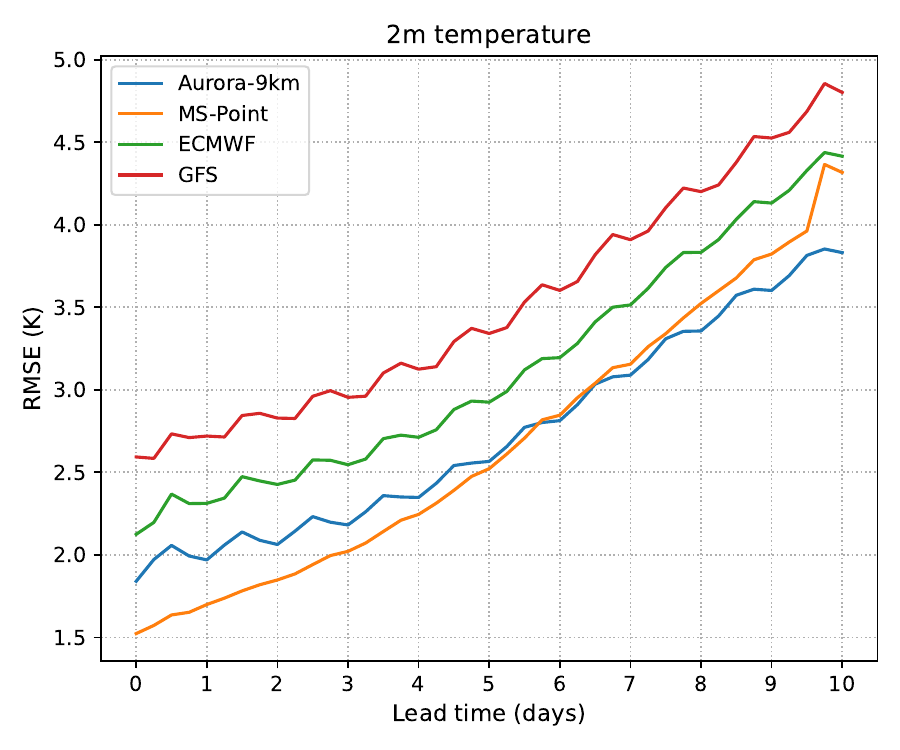}
    \includegraphics[width=0.33\textwidth]{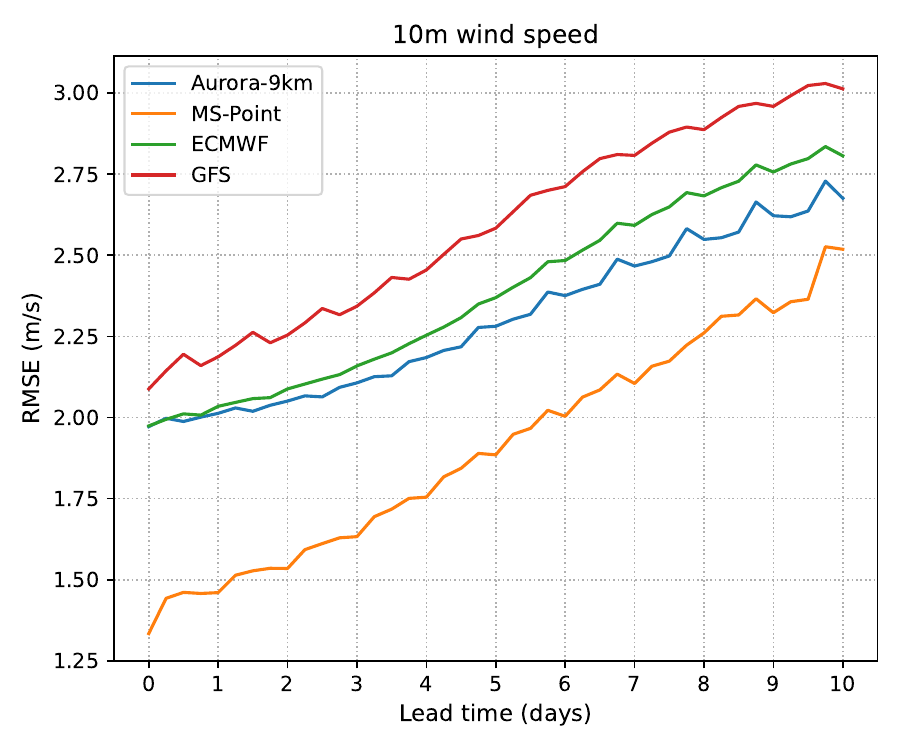} \\
    \includegraphics[width=0.33\textwidth]{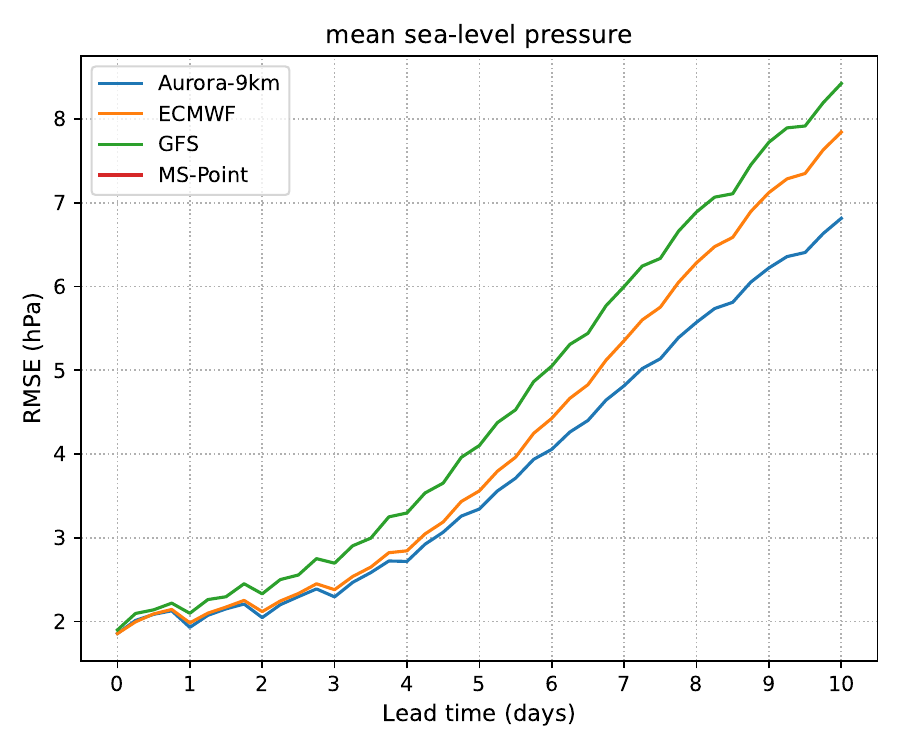}
    \includegraphics[width=0.33\textwidth]{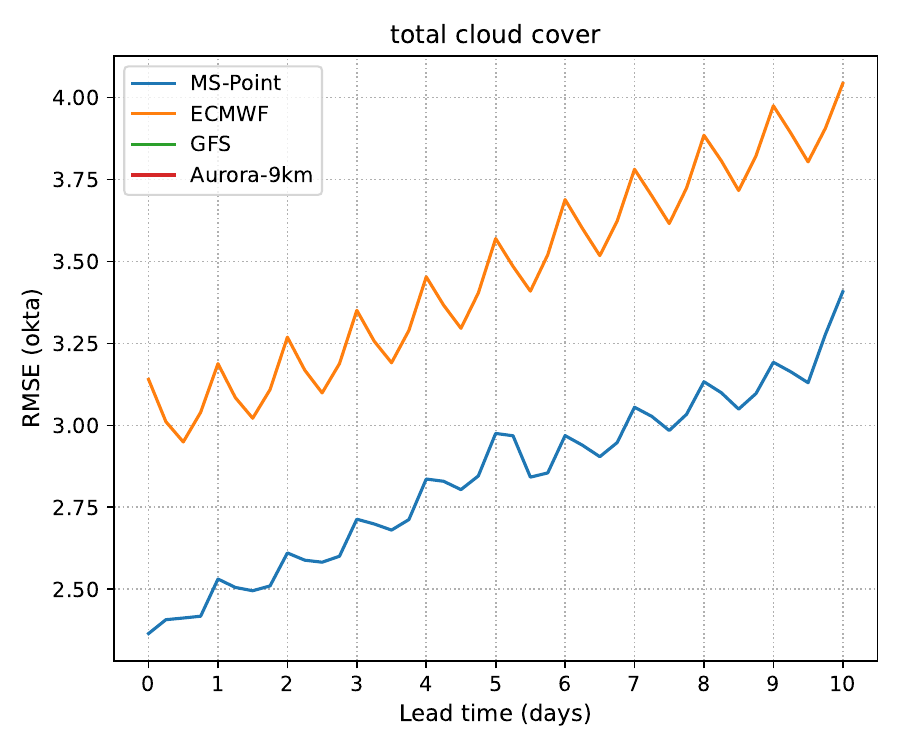}
    \caption{RMSE as a function of forecast lead time for three models evaluated against the WeatherReal-ISD observation set. Results for temperature, wind speed, mean sea-level pressure, and total cloud cover are shown.}
    \label{fig:eval_1_rmse}
\end{figure}

Figure~\ref{fig:eval_1_rmse} shows the RMSE for 2m temperature, 10m wind speed, mean sea-level pressure, and total cloud cover for forecasts up to 10 days out. We first note that the relative performance of the Aurora data-driven model against the ECMWF model follows closely with previously-reported performance against ERA5 and observation data~\citep{bodnar_aurora_2024}. This builds confidence in the results but also suggests that overall forecast accuracy is not very sensitive to the observation data that forecasts are measured against. The improvement in temperature for Aurora against ECMWF might be due in part to using the operational analysis as initial states for the model, while the IFS uses a less-accurate analysis with fewer assimilated observations. 

The flexible WeatherReal evaluation platform allows for direct comparison between grid-based and point-based forecasts, such as MS-Point. Despite being based on IFS forecasts, among other data sources, the point-based forecasts outperform gridded forecast interpolation for temperature forecasts up to 5 days out, and significantly outperform the gridded models for wind speed forecasts and cloud forecasts. We should also note that, for a limited-range variable such as cloud, the RMSE metric discourages forecasts from predicting values at the extreme ends (0 and 100\%) of the cloud fraction spectrum. Thus the WeatherReal evaluation supports using other metrics such as the equitable threat score (ETS, higher is better) with a configurable threshold (for example, 90\%, to evaluate overcast forecasts).

\begin{figure}
    \centering
    \includegraphics[width=0.49\textwidth]{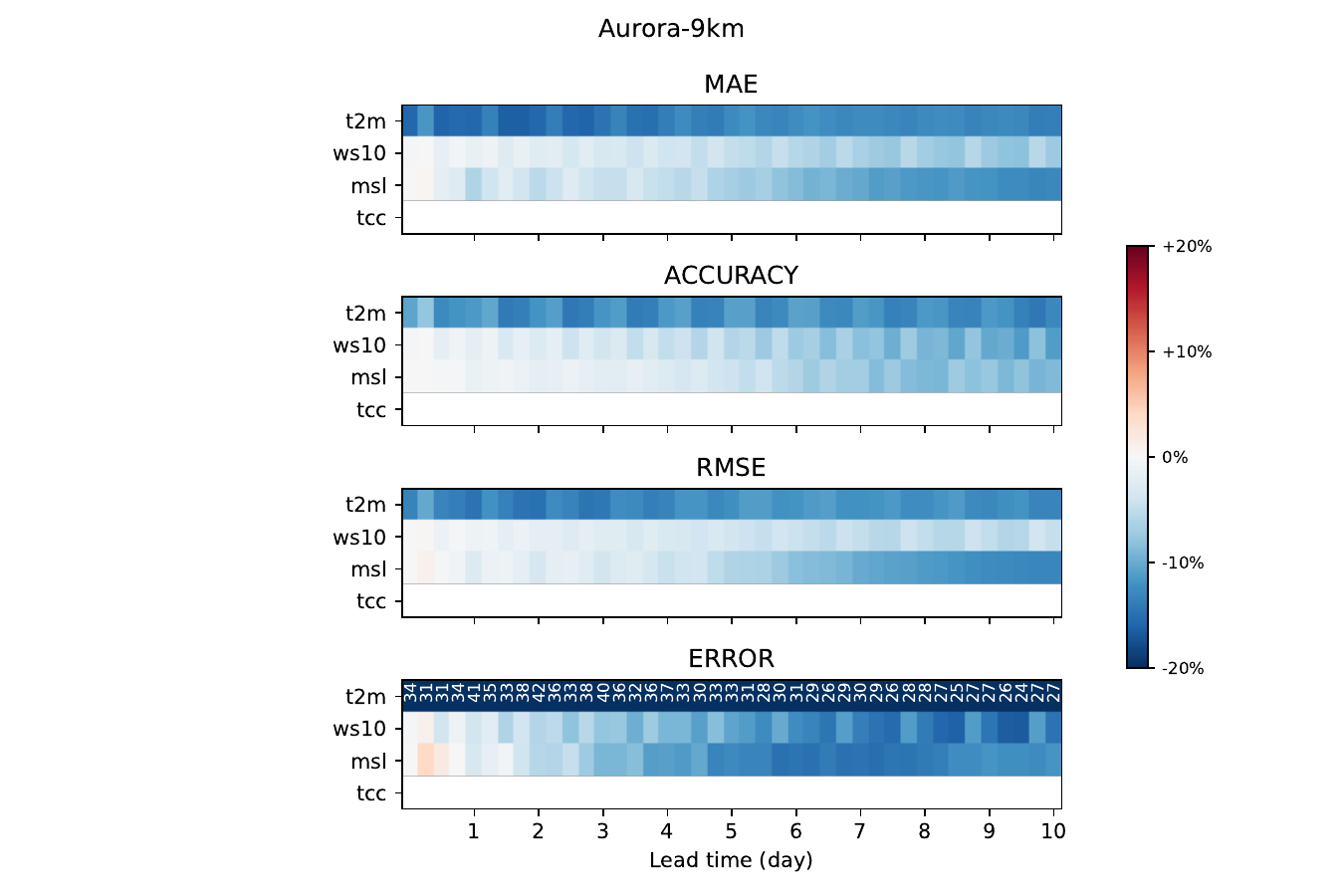}
    \includegraphics[width=0.49\textwidth]{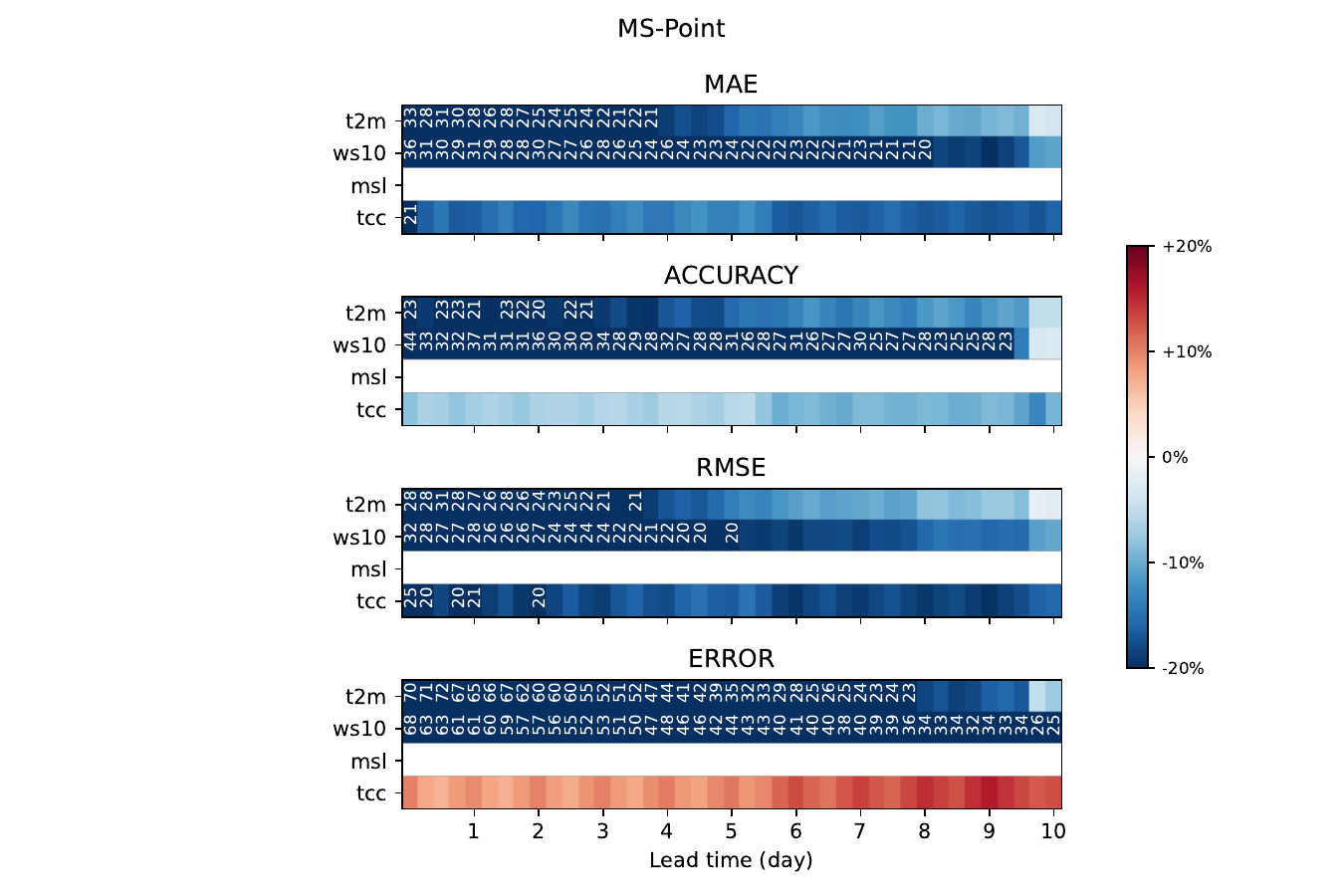}
    \caption{Scorecards for Aurora-9km (left) and MS-Point (right) forecasts evaluated against the ECMWF IFS baseline using WeatherReal-ISD observations. Colored cells indicate the percentage of metric improvement (blue) or degradation (red) versus the baseline. Values outside of the -40\% to 40\% range in the colorbar are written.}
    \label{fig:eval_1_sc}
\end{figure}

Figure~\ref{fig:eval_1_sc} shows scorecards for the Aurora-9km and MS-Point models to highlight the relative difference on several metrics supported by WeatherReal for each variable, as compared to the ECMWF baseline. In addition to RMSE and mean absolute error (MAE), we also include accuracy (defined as fraction of occurrence within a specified threshold of truth) and large error (ERROR; defined as a fraction of occurrences of forecast differing from truth by more than a specified threshold). The thresholds for accuracy and large errors are the following: for temperature, 1.7~K and 5.6~K; for wind speed, 1~m/s and 3~m/s; for mean sea-level pressure, 2~hPa and 5~hPa; and for total cloud cover, 2~oktas and 5~oktas. Generally, across these selected metrics, there isn't much observable difference in forecast performance. While the MS-Point forecast generally has decreasing margins relative the the baseline at longer lead times, Aurora-9km maintains or slightly improves its relative performance over lead time. This is likely because the data-driven model is trained with regression losses and fine-tuned over multiple time steps, which tends to lead to blurrier and more ensemble-mean-like forecasts.

\begin{figure}[htbp]
    \centering
    \includegraphics[width=\linewidth]{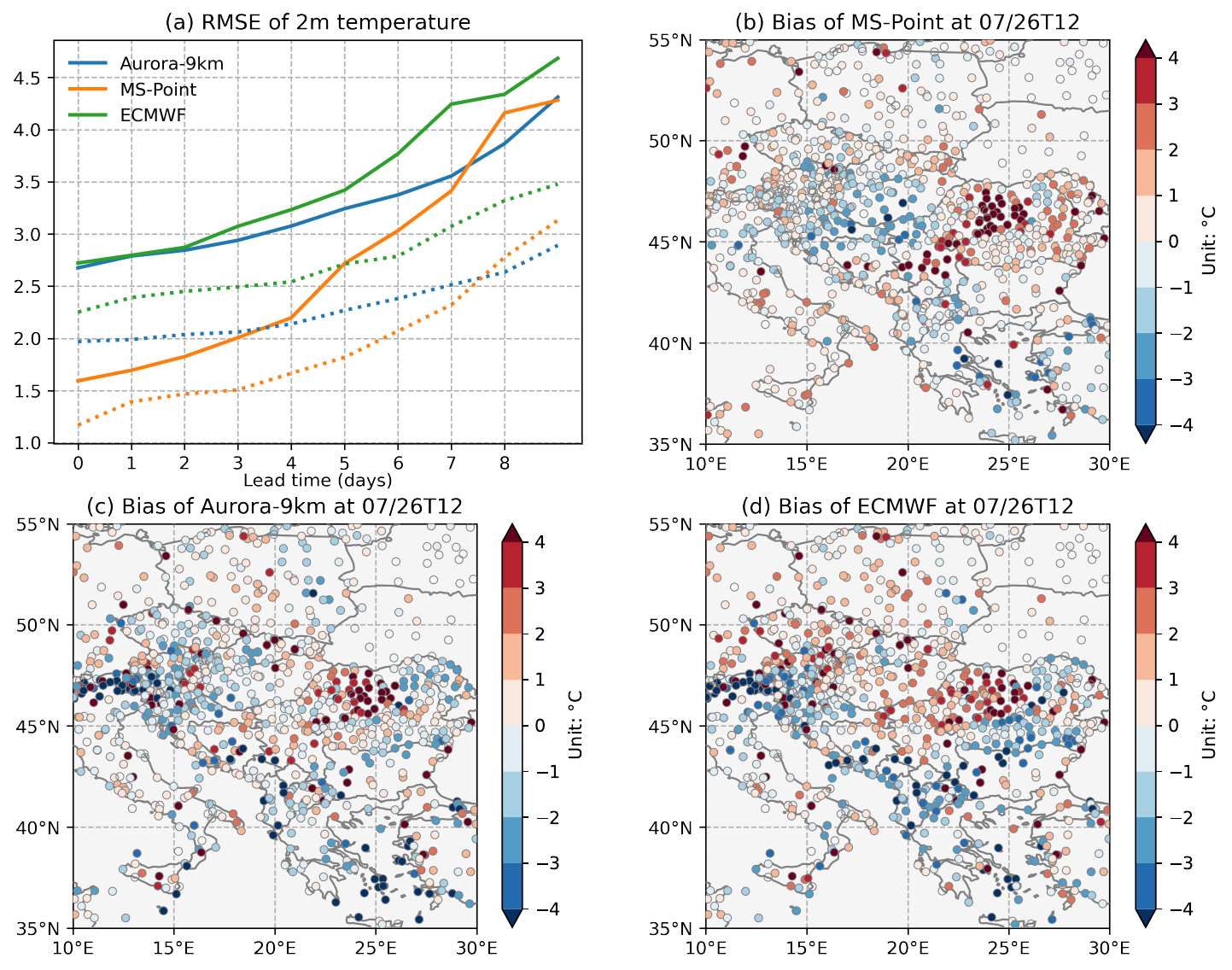}
    \caption{(a) RMSE of the 2m temperature as a function of forecast lead time for three models evaluated against WeatherReal-ISD in the region 10-30°E, 35-55°N in July 2023. The solid line represents the target time of 12 UTC, while the dashed line represents 00 UTC. (b-d) Bias of various forecasts at 12 UTC on July 26 compared to WeatherReal-ISD, with a lead time of 12 hours.}
    \label{fig:case_eval_heatwave}
\end{figure}

Evaluating forecasts for a specific period within a confined region provides insight into a model's capability to predict extreme events. Here we further evaluated these models for the region in 10-30°E, 35-55°N, as depicted in Figure 12, during July 2023. Specifically, RMSE is computed for model forecasts at 12UTC (local afternoon) and 00UTC (local early morning) (Figure~\ref{fig:case_eval_heatwave}a). Although Aurora-9km still generally outperforms the ECMWF forecast, it is important to note that for 12UTC (solid line), close to the time of the local daily maximum temperature, this advantage nearly vanishes when the lead time is within 2 days. In contrast, the MS-Point Forecast, trained on station observations, continues to hold a significant advantage. For 00UTC forecasts (dashed line), Aurora’s RMSE consistently remains lower than that of ECMWF across different lead times. Figure~\ref{fig:case_eval_heatwave}b-d further present the biases of different forecasts compared to station observations at 12 UTC on July 26. Aurora-9km and ECMWF exhibit similar patterns, which are also closely matching the bias pattern of ERA5 (Figure ~\ref{fig:case_heatwave}d).

The result observed here may originate from Aurora-9km being initialized with operational analysis from the IFS, or from Aurora-9km being primarily trained using datasets like ERA5, leading to a failure to capture the local climate characteristics of the region, such as underestimating the high temperatures induced by afternoon solar radiation in southern Europe. This underscores the importance of using in-situ observational data to train and evaluate data-driven models.

\subsection{WeatherReal-Synoptic}

\begin{figure}
    \centering
    \includegraphics[width=0.33\textwidth]{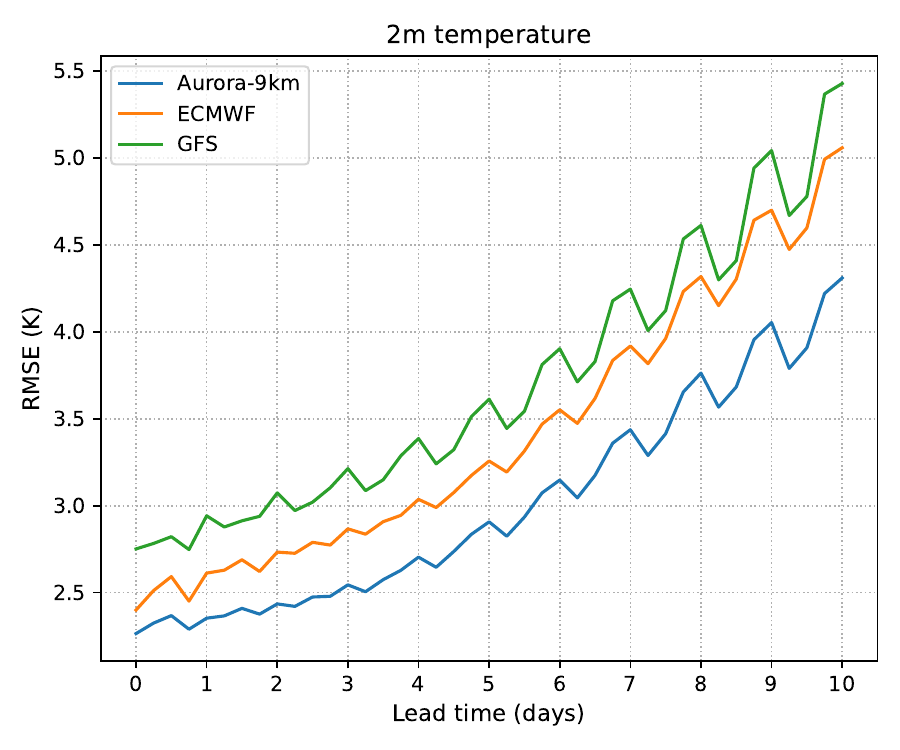}
    \includegraphics[width=0.33\textwidth]{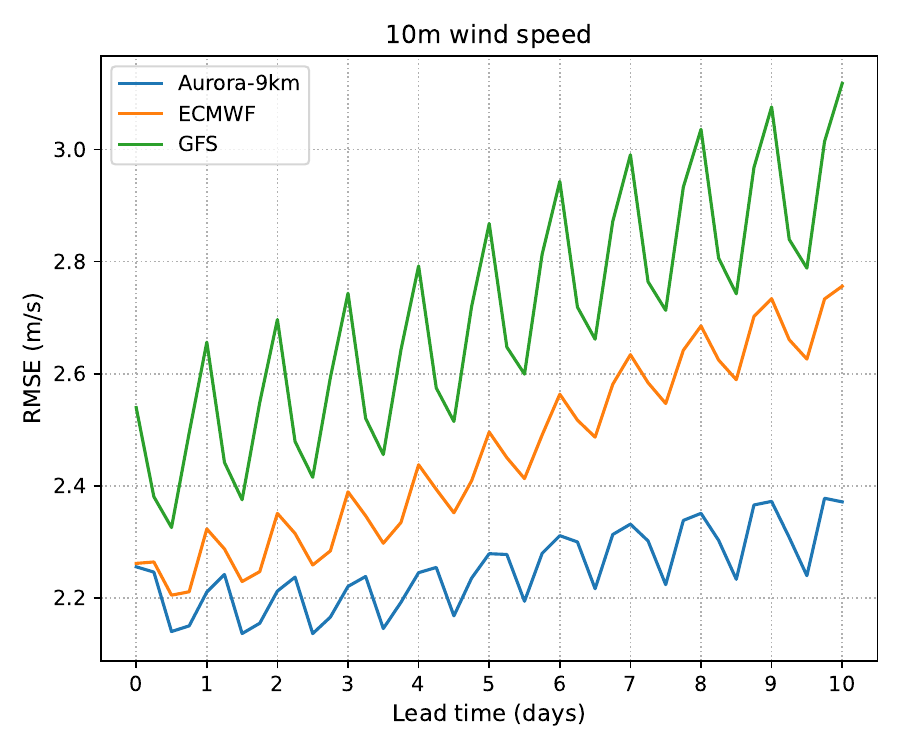}
    \includegraphics[width=0.33\textwidth]{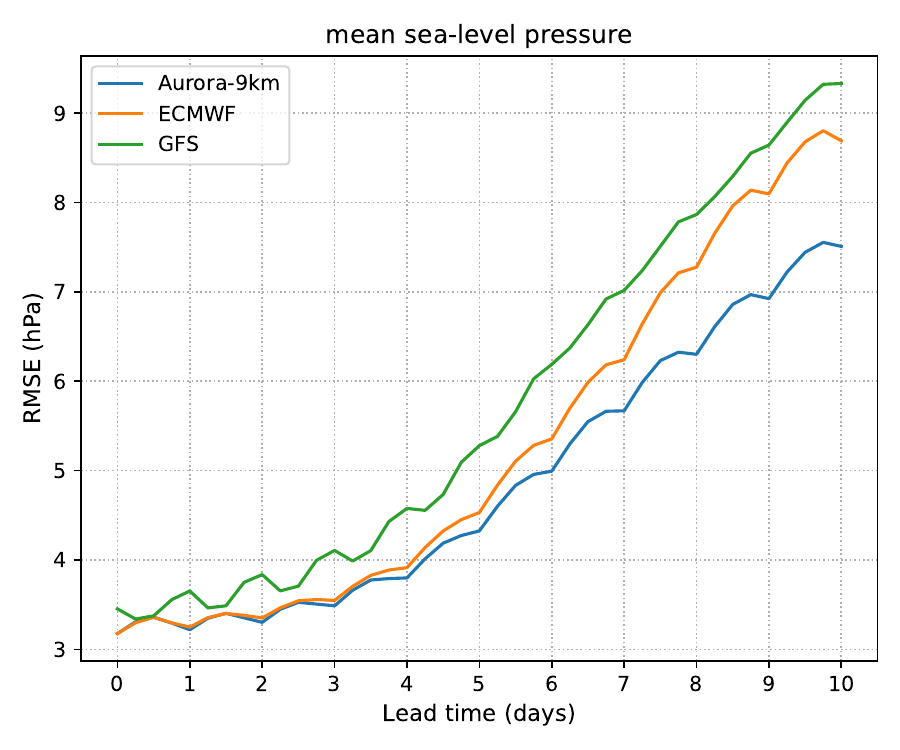}
    \caption{As in Fig.~\ref{fig:eval_1_rmse}, except the forecasts are evaluated against the WeatherReal-Synoptic observations, and omitting the cloud variable.}
    \label{fig:eval_2_rmse}
\end{figure}

\begin{figure}
    \centering
    \includegraphics[width=0.49\textwidth]{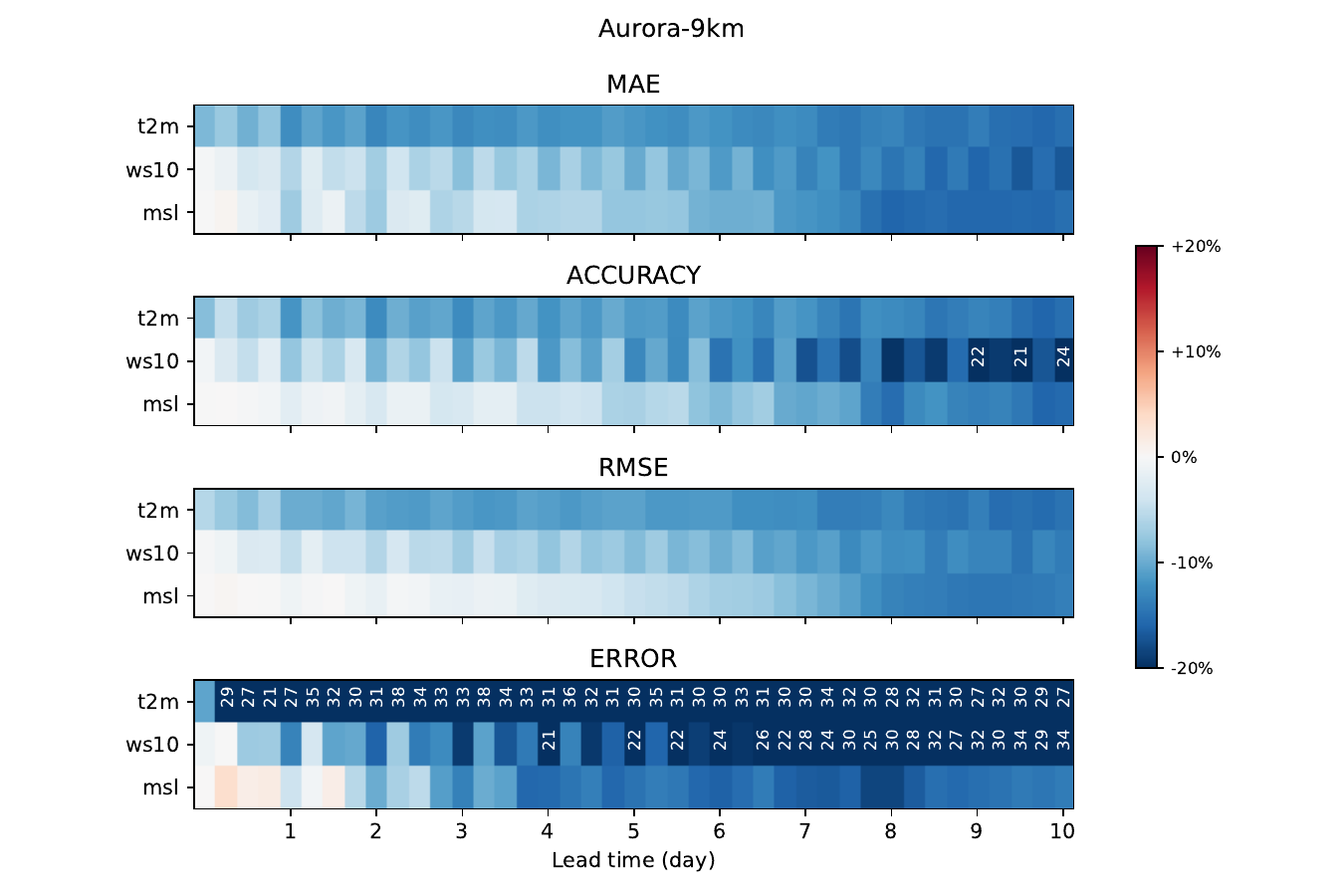}
    \caption{Scorecards for Aurora-9km forecasts evaluated against the ECMWF IFS baseline using the WeatherReal-Synoptic observation set. Other details as in Fig.~\ref{fig:eval_1_sc}.}
    \label{fig:eval_2_sc}
\end{figure}

To further explore the behavior of the models, we now run the same evaluation against the set of observations from Synoptic Data PBC (described in Section ~\ref{section:synoptic}). Figures ~\ref{fig:eval_2_rmse} and ~\ref{fig:eval_2_sc} show the same evaluations as before except that now the observation set is WeatherReal-Synoptic. MS-Point forecasts are omitted because there were insufficient point location forecasts within acceptable distances of the locations of the WeatherReal-Synoptic stations. Overall, the results are very similar across the models compared to those on the WeatherReal-ISD set: Aurora-9km maintains an impressive lead over both NWP models, with the exception being parity on mean sea-level pressure forecasts up to about day 4. The temperature gap between Aurora and ECMWF across all lead times remains despite being smaller at early lead times, an encouraging sign that the model is performing well even without the advantage of more assimilated observations in the analysis initial conditions (it is likely that most of these observations are not routinely fitted to in the IFS data assimilation process).

Also noticeable in these results is an increase in noise, particularly for wind speed errors, which have a strong diurnal cycle. This is likely due to the heavy bias of this dataset towards the United States region, with solar heating resulting in stronger afternoon/evening winds that are harder for the models to capture. For all three parameters, the errors are noticeably higher compared to those on the WeatherReal-ISD set -- likely a reflection of 1) less model generalization to regions with more sparse observations available in the data assimilation process and 2) increased noise in the data due to lower sensor and station siting quality. Nevertheless, the results are indicative of what users might observe in localized micro-climates.

\section{Tasks and Leaderboards}

\subsection{Task Definitions}

By virtue of being raw, in-situ observations, the WeatherReal dataset can be applied to many atmospheric forecasting tasks, with the medium-range forecast evaluations of Section~\ref{section:evaluation} being only one example. Since models often specialize in one aspect of atmospheric prediction (for example, short-term precipitation), we propose a few possible tasks with specific scoring methods that can be ranked across research models. These proposals are not set in stone; we welcome feedback from the modeling community on how best to use the data for evaluating forecasts in a way that best reflects the end consumer experience with various forecasting models. We propose the following:

\begin{enumerate}
    \item \textbf{Medium-range forecasting.} Forecasts are initialized twice daily at 00 and 12 UTC over the entire evaluation year 2023. Forecasts are evaluated at every 6 hours of lead time up to 168 hours (7 days). The headline metric is the RMSE for each predicted variable (except for ETS for precipitation) averaged over all forecasts and lead times.
    \item \textbf{Short-range forecasting.} Forecasts are initialized four times daily at 00, 06, 12, and 18 UTC. Forecasts are evaluated every 1 hour of lead time up to 72 hours. Headline metric is the RMSE (ETS for precipitation) for each predicted variable averaged over all forecasts and lead times.
    \item \textbf{Nowcasting.} Forecasts are initialized every hour. Forecasts are evaluated every 1 hour of lead time up to 24 hours. Headline metric is the RMSE (ETS for precipitation) for each predicted variable averaged over all forecasts and lead times.
    \item \textbf{Sub-seasonal-to-seasonal forecasts.} Following the schedule of ECMWF's long-range forecasts prior to June 2023, forecasts are initialized twice weekly at 00 UTC on Mondays and Thursdays. Forecasts are averaged either daily or weekly for lead times every 6 hours. Ideally forecasts should be probabilistic, enabling the use of proper scoring methods such as the continuous ranked probability score (CRPS). Headline metrics are week 3-4 and week 5-6 average scores.
\end{enumerate}

\subsection{Medium-Range Forecasting Provisional Leaderboard}

As noted in Section~\ref{section:evaluation}, the results shown herein are provisional since complete forecasts over the entire year of 2023 were not available for all forecast models.  Nevertheless, the current leaderboard on the medium-range forecasting task is presented in Table~\ref{tab:mr_leader}. As expected, MS-Point leads in available metrics, while Aurora-9km outperforms ECMWF where available.

\begin{table}[htbp]
\caption{Provisional leaderboard for the medium-range forecasting task, evaluated against the WeatherReal-ISD dataset}
\begin{tabular}{l p{0.15\textwidth} p{0.15\textwidth} p{0.15\textwidth} p{0.15\textwidth} p{0.15\textwidth} }
\toprule
{} &  2-m temperature &  10-m wind speed &  mean sea-level pressure &  total cloud cover &  6-hour precipitation > 1 mm \\
{} & (RMSE, K) & (RMSE, m/s) & (RMSE, hPa) & (RMSE, okta) & (ETS) \\
\midrule
MS-Point   &                      \textbf{2.258} &                        \textbf{1.753} &                                  - &                           \textbf{2.723} &                                - \\
Aurora-9km &                      2.417 &                        2.186 &                                \textbf{2.939} &                             - &                                - \\
ECMWF      &                      2.766 &                        2.251 &                                3.098 &                           3.319 &                              0.248 \\
GFS        &                      3.168 &                        2.455 &                                3.480 &                             - &                                - \\
\bottomrule
\end{tabular}
\label{tab:mr_leader}
\end{table}

\section{Discussion}

\subsection{Advantages of WeatherReal}

By incorporating in-situ observations, WeatherReal emphasizes the direct applicability of weather models in operational forecasting. Currently, most data-driven weather forecasting models adhere to the WeatherBench2~\citep{rasp_weatherbench_2024} specifications, using ERA5 for training and testing. This paper conducts a comprehensive comparison between station observations and ERA5, highlighting the limitations of reanalysis data as training and testing samples through both overall comparisons and detailed case analyses, while in-situ data provides better representation of local weather characteristics. Moreover, WeatherReal includes variables such as total cloud cover and precipitation, which are often overlooked by most current data-driven models, partly due to the intrinsic limitations of reanalysis data. These variables are critical to user needs. By evaluating mainstream NWPs and data-driven models, we demonstrate the potential application value of WeatherReal in weather model evaluation. Notably, through more detailed analysis, this dataset has a broader range of applications, including evaluation of forecasting capabilities for extreme weather events.

A rigorous quality control process is also implemented and applied to WeatherReal-ISD. Variables were extracted from the original data files, post-processed, and suspected overlapping stations in close proximity were merged. And multiple quality control algorithms were applied, resulting in the removal of about 1\% of the data to prevent outliers from impacting the evaluation results. These results have been manually verified through several rounds. Notably, the 2m temperature, 2m dewpoint temperature, mean sea-level pressure, and surface pressure have undergone extensive quality control procedures because they generally follow a normal distribution and show higher comparability with the ERA5 dataset, making them the most reliable. Conversely, the 10m wind, total cloud cover, and precipitation variables have only undergone basic quality control. Nevertheless, as station observation directly measure parameters without the involvement of any model or inversion process, they retain unique value.

\subsection{Limitations and Future Endeavors}

The current version of WeatherReal is only the initial step in our ongoing process. While WeatherReal provides observations from tens of thousands of locations worldwide, their distribution is uneven, with regions such as Africa and South America still having relatively sparse coverage. In the future, incorporating data from additional sources to achieve a denser station distribution will enhance the accuracy of evaluation results. User reports from MSN Weather are a noteworthy data source. As the number of users grows and as the user report collection system improves, we will accumulate vast amounts of real-time feedback from users about weather conditions. In this paper, user-reported temperature and precipitation has already demonstrated its value. This dataset has been used by our team to assess and improve forecast results. We plan to release this dataset at an appropriate time in the future, and hope it will provide the research community with a valuable resource for further exploration, fostering advancements in more accurate and personalized weather forecasting.

Furthermore, the quality control system described in this paper is based on statistical methods. Despite efforts to minimize manually set parameters, some erroneous records may still be missed, and some correct records may be mistakenly removed. We will continuously refine our quality control algorithms and explore the use of machine learning methods to enhance the accuracy of quality control.

Finally, we have initially released datasets for 2023 only. In-situ observation datasets are not only useful for evaluating weather models but also play a critical role in data assimilation and end-to-end models based on data-driven methods. For such a dataset to prove useful in improving data assimilation products, and subsequently training data for the latest AI weather models, a much longer record is necessary. Additionally, to more comprehensively characterize the state of the atmosphere at near-surface and various upper levels, there is potential to incorporate data from satellites, radar, and weather balloons into the WeatherReal dataset. These additional data sources may provide a more detailed and multi-dimensional understanding of atmospheric conditions.

\section{Conclusion}

We presented the WeatherReal benchmark, designed for the evaluation of weather models, with a particular focus on the latest data-driven forecasting models, based on in-situ observation data. WeatherReal currently consists of two subsets derived from station observations: one from ISD and the other from Synoptic Data PBC. Additionally, a third dataset based on user reports is under development and will be released in the future. Instructions on how to obtain the datasets and evaluation code will be provided on GitHub at \url{https://github.com/microsoft/WeatherReal-Benchmark}.

The goal of WeatherReal is to evaluate weather models using reliable, comprehensive datasets and a unified, stable framework, emphasizing that optimizing models towards in-situ observations is a fundamental prerequisite for their application in real, user-impacting operational forecasts. This ensures accurate weather forecasts and effectively addresses real-world forecasting challenges. We believe that promoting the use of in-situ observations as evaluation targets is a crucial step in the development of data-driven weather models.

\section{Acknowledgements}

We sincerely thank NOAA and ECMWF for their efforts in constructing datasets, and for their open sharing of these data. Their work has been the primary sources for WeatherReal and this paper. We also extend our gratitude to Synoptic Data PBC for aggregating data from numerous sources. Their products and services empower public, private, government, and academic users with real-time and historical data, ensuring they can make informed decisions swiftly and confidently. Finally, we also thank Cristian Bodnar and the Aurora team for providing inference results from their model for evaluation.

\bibliographystyle{unsrtnat}
\bibliography{references}

\end{document}